\documentclass[11pt,a4paper]{article}
\usepackage{amsfonts}
\usepackage[dvips]{graphicx}
\usepackage{amsthm}
\usepackage{amsmath}
\usepackage{epsfig}
\usepackage{t1enc}
\usepackage{amssymb}
\usepackage{latexsym}
\usepackage{bm}
\usepackage[hang]{caption}

\def\r{\rho} \def\d{\delta} \def\D{\Delta} \def\m{\mu}
\def\s{\sigma} \def\l{\lambda} \def\sp{{(S)}}
  \def\bl{\bar \l}
\def\H{{\cal H}}  \def\hH{\tilde H}
\def\E{{\cal E}} \def\B{{\cal B}} 
\def\a{\alpha} \def\b{\beta}
\def\tw{\tilde w} \def\s{\sigma} \def\S{\Sigma}
\def\hc{\hat c} \def\C{{\cal C}}
\def\g{\gamma} \def\G{\Gamma} \def\z{\zeta}
\def\P{{\cal P}} \def\Q{{\cal Q}} \def\R{{\cal R}}
 \def\0{{(0)}} \def\i{{id}}
\def\Bl{\Big(} \def\Br{\Big)} \def\BL{\Big[} \def\BR{\Big]}
\def\Pz{\Phi_0} \def\e{\epsilon} \def\td{\tilde \d}
\def\tim{\mbox{\scriptsize $\ \times\ $}}
\def\tP{\tilde \Phi} \def\tPs{\tilde \Psi} \def\tE{\tilde \E}
\def\te{\tilde \eta} \def\tte{\tilde{\te}}
\def\tz{\tilde z} \def\ttz{\tilde{\tz}} \def\I{{II}}
\def\tx{\tilde x} \def\ttx{\tilde{\tx}}
\def\ty{\tilde y} \def\tty{\tilde{\ty}}
\def\tX{\tilde X} \def\tY{\tilde Y} 
\def\hM{{\hat M}} \def\hN{{\hat N}} \def\1{{(1)}}

\title{\bf Cosmological perturbations \\ in the presence of a solid
with positive pressure}

\author{Vladim\'ir Balek\footnote{e-mail
address: balek@fmph.uniba.sk}\ \ and Matej \v
Skovran\footnote{e-mail address: skovran@fmph.uniba.sk}
\\
{\it Department of Theoretical Physics, Comenius University,
Bratislava, Slovakia}}

\begin{document}

\renewcommand{\figurename}{Fig.}

\maketitle
\maketitle\abstract

{Evolution of scalar perturbations in a universe containing solid
matter with positive pressure is studied. Solution for pure solid
is found and matched with solution for ideal fluid, including the
case when the pressure to energy density ratio $w$ has a jump. Two
classes of solutions are explored in detail, solutions with
radiation-like solid ($w = 1/3$) and solutions with stiff solid
($w > 1/3$) appearing in a universe filled with radiation. For
radiation-like solid, an almost flat spectrum of large-scale
perturbations is obtained only if the shear stress to energy
density ratio $\xi$ is close to zero, $|\xi| \lesssim 10^{-5}$.
For a solid with stiff equation of state, large-scale
perturbations are enhanced for $\xi$ negative and suppressed for
$\xi$ positive. If the solid dominated the dynamics of the
universe long enough, perturbations could end up suppressed as
much as by several orders of magnitude, and in order that the
inclination of the large-scale spectrum is consistent with
observations, radiation must have prevailed over the solid long
enough before recombination. In Newtonian gauge, corrections to
metric and energy density are typically much greater than 1 in the
first period after the shear stress appears, but the linearized
theory is still applicable because the corrections stay small when
one uses the proper-time comoving gauge.}


\section{Introduction}

One of possible modifications of the concordance $\Lambda$CDM
model is adding solid matter with negative parameter $w$ (pressure
to energy density ratio) to the known components of the universe
\cite{{bs},{bbs}}. Such matter can be composed of frustrated
cosmic strings, which have $w = - 1/3$, or domain walls, which
have $w = - 2/3$. The solid can, in principle, function as dark
energy, making the universe accelerate with no unstable
perturbations appearing in it. Acceleration of the universe,
however, requires $w < - 1/3$, so that it cannot be achieved with
cosmic strings. Domain walls are of no help either, because the
parameter $w$ of the acceleration driving medium must be in
absolute value greater than 2/3 to reconcile observations, as
established first for constant $w$ \cite{mel} and then for $w$
depending on time \cite{{bis},{bis1},{zha}}. Moreover, domain
walls in an expanding universe seem to be scaling rather than
frustrated \cite{{ave},{lei}}, so that their number per {\it
horizon}, rather than per unit comoving volume, is constant. This
leads to an effective $w$ which is in absolute value less than
2/3. (In a radiation dominated universe it equals $-1/3$.)
Nevertheless, scenarios containing solid matter with negative $w$
continue to attract interest. The theory was extended to
anisotropic solids \cite{{bm1},{bm2}} and used as a general
framework to describe the dark sector exclusively by means of
metric tensor \cite{{bm},{bp}}; effects of a lattice of cosmic
strings superimposed on the conventional dark sector were studied
\cite{kum}; and a possibility that solid matter could replace
inflaton field as an inflation driving medium was contemplated
\cite{{gru},{end},{bar},{sit}}.

Existing literature on cosmology with solid matter apparently does
not contain an analysis of the case with {\it positive} $w$. This
case is obviously less interesting than that with negative $w$,
which offers an alternative view of the least understood component
of the known universe. The study of solids with positive $w$ can
be nevertheless useful, if only because it would extend the
parametric space of the theory and give us greater freedom when
explaining observational data.

The concordance model gives satisfactory explanation of
observations, so that there is no need to add a new component to
it. However, it is {\it possible} to do so and not to come into
conflict with the data. The modified theory can yield the same
power spectrum of CMB anisotropies as the original one if it
assigns different values to cosmological parameters, or if it
imposes different initial conditions on cosmological
perturbations. In the former case, the theory could be verified in
the future by alternative methods of measuring cosmological
parameters, like determining $h$ from the data on double quasars
or from Zel'dovich-Sunyaev effect, if they become precise enough;
in the latter case, the modified theory would pick a different set
of inflationary scenarios than that preferred by the original
theory, and could be verified by measuring quantities like
tensor-to-scalar ratio sensitive to the details of inflation. To
demonstrate these two possibilities, consider the two specific
kinds of solids explored in the text. For a universe with both
radiation-like and stiff solid, the theory predicts different
behavior of large-scale anisotropies than for a universe with pure
fluid, but this cannot be cured by changing the parameters of the
concordance model; instead, we must tune the new parameters
entering the theory. However, a radiation-like solid could
influence significantly also the behavior of {\it short-scale}
anisotropies, provided it has viscosity and thus produces an
additional Silk dumping. The total dumping could be then
suppressed by reducing the value of $\Omega_m h^2$. On the other
hand, after we ensure that large-scale anisotropies are the same
on all scales, the presence of a stiff solid still affects their
{\it size}, provided the initial size of cosmological
perturbations is fixed. If the shear modulus is positive, the
perturbations are {\it suppressed}, yielding smaller anisotropies
than in a universe with pure fluid; thus, to accommodate
observations the initial perturbations must be greater.

In order that a solid plays any role in the evolution of the
universe, it must stay solid while being stretched by many orders
of magnitude. ``Stay solid'' means that its shear modulus, or
moduli if it is anisotropic, are nonzero and comparable with the
pressure acting inside it. Solids with positive pressure can have
effect on the dynamics of the universe during the first, radiation
dominated era, only if their parameter $w$ satisfies $w \ge 1/3$.
Thus, their pressure must be comparable with their energy density,
and if we want that they are significantly distinct from fluid,
their shear modulus must be comparable with their energy density,
too.

What kind of matter could have desired properties? One possibility
is Coulomb crystal, provided its mobile charges have high enough
energies. Coulomb crystals composed of ions with one missing
electron and free electrons, both with particle number density
$n$, are known to have shear modulus of order $e^2 n^{4/3}$
\cite{fu}. The theory applies also to the highly compressed matter
inside white dwarfs and neutron stars crusts, after it has
sufficiently cooled down \cite{{hor},{bai}}. In this case the
standing charges are nuclei rather than ions, therefore the
previous expression modifies to $Z^2 e^2 n_0^{4/3}$, where $n_0$
is the density of nuclei and $Z$ is their atomic number. Consider
a crystal consisting of two kinds of {\it nonelectrically} charged
particles, movable ones with the charge $-g$ and standing ones
with the charge $Zg$. (Electrically charged particles, when put
into a hot universe, would not form a crystal but a gas of
particle-antiparticle pairs with the same temperature as that of
radiation.) If $Z \gg 1$, the movable particles can have Fermi
energy comparable with the rest mass of the later particles, or
even much greater than it, and the standing particles can be still
separated by distances greater than their Compton wavelength. Such
matter would have energy density of order $n^{4/3}$, where $n =
Zn_0$ is the density of the movable particles. The parameter $w$
of the matter would be 1/3; and the shear modulus would differ
from the energy density by the factor $Z^{2/3} g^2$, which might
well be of order 1 or greater.

A question arises whether a compressed system of {\it equally}
charged particles cannot be a solid, too. Such system with
short-range forces between the particles was considered as a
candidate for the matter in the center of neutron stars
\cite{{zel},{wal}}. If the particle density is large enough,
energy density of the system reduces to its potential energy per
unit volume and is of order $g^2 m^{-2} n^2$, where $g$ is the
charge of the particles and $m$ is the mass of the Yukawa field
mediating the interaction between the particles. This corresponds
to $w = 1$ (``extremely stiff equation of state''). It could seem
that if the particles are arranged into a lattice, after being
shifted, they would return to their original positions due to the
repulsive forces from their neighbors. If this was the case, the
system would be solid rather than fluid. However, it can be shown
by direct calculation that the shear modulus of the lattice is
zero. This can be understood in terms of the stress tensor of the
Yukawa field: with the $1/r$ singularities removed from the field
(since the corresponding stress is compensated by the forces
holding the charges together), the tensor is obviously isotropic.

One can also speculate about another kind of solid with positive
pressure, a lattice of strings or walls whose energy is {\it
inversely} proportional to their length (in case of strings) or
area (in case of walls). Such objects can be called
``springlike'', since they behave as a compressed spring. Denote
the shear modulus to energy density ratio of the solid by $\xi$.
Lattices of ordinary strings and walls, both random and ordered,
were shown to have $\xi$ positive and large enough to make the
longitudinal sound speed squared positive \cite{{bccm},{bcm}}.
(The cubic lattice considered in \cite{bcm} has {\it two} $\xi$'s,
longitudinal and transversal, defining the corresponding sound
speeds. For such lattice, the claim is that both $\xi$'s are
positive and the longitudinal one yields positive sound speed
squared.) Springlike strings and walls, on the other hand, have
{\it negative} $\xi$; in particular, when arranged into a random
lattice they have $\xi = -2/15$. However, their parameter $w$ is
positive, $w = 1/3$ for strings and $w = 2/3$ for walls, and this
suffices for the longitudinal sound speed squared to be positive.

Negative $\xi$ seems to be excluded since the transversal sound
waves in a solid with such $\xi$ are unstable (their velocity
squared is negative). Surprisingly, this might not be as
devastating for a solid in the early universe as it would be for a
solid in laboratory. Vector perturbations (a different name for
transversal sound waves) are usually not considered in cosmology
because they fall down as (scale parameter)$^{-2}$ \cite{mukh}.
However, in the theories in which vector fields are conformally
coupled they are absent completely since they cannot be formed
during inflation \cite{par}. Thus, if we restrict ourselves to
such theories, instability of vector perturbations does not
matter. There is no exponential growth if the initial value of the
function is zero.

After the radiation passes a part of its energy to the solid,
further evolution of the perturbations depends on how the lattice
which was formed in the process looks like. A natural assumption
is that it is {\it relaxed}; that is, it looks locally the same as
when created in laboratory and stretched (for $w < 0$) or
compressed (for $w > 0$) by the same forces from all sides. It
does not follow the shear deformation of the volume elements of
the fluid (that is, radiation), since such deformation has no
effect on the distribution of the particles as long as they move
freely. Instead, the deformation of the fluid translates into the
{\it internal geometry} of the solid (its body geometry in the
relaxed state) \cite{bs}. Obviously, such isotropic solidification
means that no shear stress acts in the solid at the moment it is
formed. This lasts until the perturbation starts to evolve, since
there is no shear stress without shear deformation. As a result,
in a solid with $w > 0$ the presence of $\xi$ shows up only after
the perturbation enters the horizon. Long-wavelength perturbations
are not affected at all. There is only one way to make these
perturbations sensitive to $\xi$: anisotropic solidification.
However, such process cannot take place in common fluid, whose
solidification means arranging freely moving particles into a
lattice. Anisotropy has to be latently present, inherited from the
process in which the perturbations were created. A scenario to
provide a mechanism for that which seems to be most promising is
solid inflation proposed in \cite{{gru},{end},{bar},{sit}}.

In the paper we study scalar perturbations in a universe
containing solid matter with positive pressure. In section
\ref{sec:cosmo} we solve equations governing the evolution of
perturbations for a solid with constant parameters $w$ and $\xi$
and match the solutions for a fluid and a solid with the same $w$;
in section \ref{sec:rad} we explore long-wavelength perturbations
in a universe containing radiation-like solid ($w = 1/3$); in
section \ref{sec:stiff} we investigate the same kind of
perturbations in a universe filled with radiation in which there
appears stiff solid ($w > 1/3$); and in section \ref{sec:con} we
summarize the results. Signature of the metric tensor is $(+ - - -
)$ and a system of units is used in which $c = 16\pi G = 1$.


\section{Cosmology with shear stress}
 \label{sec:cosmo}

\subsection{Description of perturbations}

Consider a flat FRWL universe filled with an elastic medium, fluid
or solid, and add a small perturbation to it. The perturbed metric
is
\begin{equation}
ds^2 = a^2 (d\eta^2 - d{\bf x}^2 + h_{\mu \nu} dx^\mu dx^\nu),
 \label{eq:ds2}
\end{equation}
where $\eta$ is conformal time and $a$ is scale parameter. In an
unperturbed universe, the medium is characterized by the mass
density $\r$ and pressure $p$. In a perturbed universe, these
quantities acquire small corrections $\d \r$ and $\d p$ and the
medium itself acquires small velocity ${\bf v} \doteq a{\bf u}$,
where ${\bf u}$ is the 3-space part of the 4-velocity $u^\mu =
dx^\mu/d\tau$. For an ideal fluid, the perturbed stress-energy
tensor can be expressed in terms of $\d \r$, $\d p$ and $\bf u$ as
\begin{equation}
{T_0}^0 = \r + \d \r, \quad {T_i}^0 = a^{-1} \r_+ u_i, \quad
{T_i}^j = - (p + \d p) \d_{ij},
 \label{eq:Tf}
\end{equation}
where $\r_+ = \r + p$ and $u_i = - a^2 u^i + a h_{0i}$. For a
solid, an additional term $\D {T_i}^j$ coming from shear stress
appears in ${T_i}^j$.

Expressions for $\d \r$, $\d p$ and $\D {T_i}^j$ can be obtained
from the theory of relasticity (relativistic elasticity),
summarized in appendix \ref{app:rel}. Suppose the solid is
homogenous and isotropic, so that its elastic properties are
completely described by the Lame coefficients $\l$ and $\mu$. The
combination of the two coefficients $K = \l + 2\mu/3$ is {\it
compressional modulus} and the coefficient $\mu$ is {\it shear
modulus}. (Our $K$ is 2 times greater and our $\mu$ is 4 times
greater than $K$ and $\mu$ in \cite{bs}. We have defined them so
in order to be consistent with the standard definitions in
Newtonian elasticity.) Suppose, furthermore, that the solid has
Euclidean internal geometry and no entropy perturbations are
present in it, so that the perturbation to the stress-energy
tensor is given exclusively by the perturbation to the 3-metric
$h_{ij}$ and by the shift vector $\bm \xi$. Denote the stress
tensor and its perturbation by $\tau_{ij}$ and $\d \tau_{ij}$,
$\tau_{ij} = - p \d_{ij}$ and $\d \tau_{ij} = - \d p \d_{ij} + \D
{T_i}^j$. In the comoving gauge $\bm {\xi} = 0$ and the functions
$\d \r$ and $\d \tau_{ij}$ are of the form
\begin{equation}
\d \r = \frac 12 \r_+ h_{kk}, \quad \d \tau_{ij} = - \frac 12 \l
h_{kk} \d_{ij} - \mu h_{ij}.
 \label{eq:dcom}
\end{equation}
The velocity $\bf v$ can be written in terms of the shift vector
as ${\bf v} = {\bm \xi}'$, where the prime denotes differentiation
with respect to $\eta$; thus, in the comoving gauge $\bf v$ as
well as $\bf u$ vanish and $u_i = a h_{0i}$.

Evolution of an unperturbed universe is described by the equations
\begin{equation}
a' = \Bl\frac 16 \r a^4\Br^{1/2}, \quad \r' = - 3\H\r_+,
 \label{eq:Eeq}
\end{equation}
where $\H$ is {\it Hubble parameter} with respect to conformal
time, $\H = a'/a$. Note that the equation for $\r$ follows from
the first equation in (\ref{eq:id0}) if we insert $V \propto a^3$
into it. From the second equation in (\ref{eq:id0}) we obtain in
the same way
\begin{equation}
p' = -3 \H K.
 \label{eq:id}
\end{equation}

We will be interested in {\it scalar perturbations} only. These
perturbations are most easily interpreted in the {\it Newtonian
gauge}, in which the scalar part of the metric is
\begin{equation*}
ds^{\sp 2} = a^2 [(1 + 2\Phi)d\eta^2  - (1 - 2\Psi) d{\bf x}^2].
\end{equation*}
Thus, in the Newtonian gauge scalar perturbations to the metric
are described solely by two functions, Newtonian potential $\Phi$
and an additional potential describing the curvature of 3-space
$\Psi$. Write the scalar part of the 3-tensor $\d \tau_{ij}$ as
\begin{equation*}
\d \tau_{ij}^\sp = \tau^{(1)} \d_{ij} + {\tau^{(2)}}_{\mbox{\hskip
-1mm},ij}.
\end{equation*}
Einstein equations yield three equations for five functions
$\Phi$, $\Psi$, $\d \r$, $\tau^{(1)}$ and $\tau^{(2)}$
\cite{mukh}. We will use just one of them, the algebraical
constraint
\begin{equation}
\Phi = \Psi + \frac 12\tau^{(2)} a^2.
 \label{eq:diff}
\end{equation}
Note that an ideal fluid has $\tau^{(2)} = 0$, hence in a universe
filled with ideal fluid the potentials $\Phi$ and $\Psi$ coincide.

To find the time dependence of $\Phi$ and $\Psi$, one can
determine $h_{\mu \nu}$ as functions of $\eta$ in {\it any} gauge
and compute $\Phi$ and $\Psi$ from them. Moreover, one can define
the functions $\overline{\d \r}$, $\overline{\tau^{(1)}}$ and
$\overline{\tau^{(2)}}$ as invariant functions reducing to $\d
\r$, $\tau^{(1)}$ and $\tau^{(2)}$ in Newtonian gauge, and compute
them from the functions $\d \r$, $\tau^{(1)}$ and $\tau^{(2)}$ in
the gauge one is working in. If we write the scalar part of the
perturbed metric in a general gauge as
\begin{equation*}
ds^{\sp 2} = a^2 [(1 + 2\phi) d\eta^2 + 2 B_{,i} d\eta dx^i -
(\d_{ij} - 2\psi \d_{ij} - 2E_{,ij}) dx^i dx^j],
\end{equation*}
the expressions for gauge invariant functions are
\begin{equation}
\Phi = \phi - \D' - \H \D, \quad \Psi = \psi + \H \D, \quad
\overline{\d \r} = \d \r - \r' \D, \quad \overline{\tau^{(1)}} =
\tau^{(1)} + p' \D,
 \label{eq:db}
\end{equation}
where $\D = B - E'$ \cite{mukh}. The function $\tau^{(2)}$ is
already gauge invariant, $\overline{\tau^{(2)}} = \tau^{(2)}$.

In our description of perturbations in the presence of solid we
will follow \cite{pol} and use the {\it proper-time comoving
gauge}, defined by the conditions $\phi = 0$ and ${\bm \xi} = 0$.
In this gauge, the cosmological time $t = \displaystyle \int a
d\eta$ is the proper time of the observers at rest and the
observers move with the matter. The gauge is not defined uniquely
since one can shift cosmological time by an arbitrary function $\d
t({\bf x})$. Under such shift, $E$ stays unaltered and $B$ and
$\psi$ transform as
\begin{equation*}
B \to B + \d \eta, \quad \psi \to \psi - \H \d \eta,
\end{equation*}
where $\d \eta = a^{-1} \d t$. This suggests that we represent $B$
and $\psi$ as
\begin{equation*}
B = \B + \chi, \quad \psi = - \H \chi,
\end{equation*}
where $\B$ stays unaltered by the time shift and $\chi$ transforms
as $\chi \to \chi + \d \eta$. As a result, we obtain an expression
for $\Psi$ that is explicitly time-shift invariant,
\begin{equation}
\Psi = \H (\B - E').
 \label{eq:Ps0}
\end{equation}
For $\Phi$ expression (\ref{eq:diff}) will be used, but one can
easily check that the expression in (\ref{eq:db}) is time-shift
invariant, too. Indeed, it contains $\chi$ only in the combination
$a^{-1}(a\chi)'$, and the product $a\chi$ gets shifted by $\d t$
which does not depend on $\eta$.

Formulas for $\d \r$, $\tau^{(1)}$ and $\tau^{(2)}$, obtained by
expressing the scalar part of $h_{ij}$ in (\ref{eq:dcom}) in terms
of $\psi$ and $E$, are
\begin{equation*}
\d \r = \r_+ (3\psi + \E), \quad \tau^{(1)} = - 3K \psi - \l \E,
\quad \tau^{(2)} = - 2\mu E,
\end{equation*}
where $\E = \triangle E$. After inserting from the first two
formulas into the definitions of $\overline{\d \r}$ and
$\overline{\tau^{(1)}}$ and using expressions (\ref{eq:Eeq}) and
(\ref{eq:id}) for $\r'$ and $p'$, we find
\begin{equation}
\overline{\d \r} = \r_+ (3\Psi + \E), \quad \overline{\tau^{(1)}}
= - 3K\Psi - \l \E.
 \label{eq:deb}
\end{equation}
Finally, equation (\ref{eq:diff}) with $\tau^{(2)}$ inserted from
the third formula reads
\begin{equation}
\Phi = \Psi - \mu a^2 E.
 \label{eq:Ph0}
\end{equation}

We will restrict ourselves to perturbations of the form of plane
waves with the wave vector $\bf k$, $\B$ and $E \propto e^{i{\bf
k} \cdot {\bf x}}$. The action of the Laplacian then reduces to
the multiplication by $- k^2$; in particular, the definition of
$\E$ becomes $\E = - k^2 E$. To simplify formulas, we will
suppress the factor $e^{i{\bf k} \cdot {\bf x}}$ in $\B$ and $E$,
as well as in the other functions describing the perturbation.
They will be regarded as functions of $\eta$ only.

For the functions $\B$ and $\E$ we have two coupled linear
differential equations of first order, coming from equations
${{T_i}^\mu}_{;\mu} = 0$ and $2G_{00} = T_{00}$. They can be
obtained from equations for $y_{01}$ and $y_{11}$ in \cite{pol} by
putting $e^z = a$, $y_{01} = a \B$, $y_{11} = -2\E$, $\e = a^3 \r$
and $\s = a^3 p$ and replacing $\l + \s \to a^3 \l$ and $\mu + \s
\to a^3 \m$. The equations are
\begin{equation}
\B' = (3c_{S0}^2 + \a - 1)\H \B + c_{S\|}^2 \E, \quad \E' = - (k^2
+ 3\a \H^2) \B - \a \H \E,
 \label{eq:dBdE}
\end{equation}
where $\a = (2\H)^{-2}\r_+ a^2$ and the sound speeds $c_{S0}$ and
$c_{S\|}$ are defined in equations (\ref{eq:cS20}) and
(\ref{eq:cS2}). Pressure and shear stress of the medium can be
characterized by the dimensionless parameters $w = p/\r$ and $\xi
= \mu/\r$; however, to simplify formulas we will use $\b =
\mu/\r_+$ instead of $\xi$. The only place where the parameter
$\b$ enters equations (\ref{eq:dBdE}) is the term $c_{S\|}^2 \E$
in the equation for $\B$, since $c_{S\|}^2 = c_{S0}^2 + (4/3)\b/$
and $c_{S0}^2$ does not contain $\b$.

In the concordance model, one does not consider shear stress but
introduces another source of anisotropic stress -- {\it
viscosity}. Let us compare the effect of viscosity with the effect
of shear stress in our gauge. If we introduce one more
dimensionless parameter $\g = {\bm \eta}/\r_+$, where $\bm \eta$
is  the coefficient of shear viscosity, the equation for $\B$ is
\begin{equation*}
\B' = (3c_{S0}^2 + \a - 1)\H \B + c_{S0}^2 \E +
 \bigg\{ \mbox{\hskip -2mm}
  \left. \begin{array} {l}
  (4/3) \b \E \mbox{ for a solid}\\
  (4/3) \g \E' \mbox{ for a viscuous fluid}\\
  \end{array}\mbox{\hskip -1mm}. \right.
\end{equation*}

For $\Phi$ and $\Psi$ we have equations (\ref{eq:Ps0}) and
(\ref{eq:Ph0}). After inserting into the former equation from the
second equation (\ref{eq:dBdE}) and into the latter equation from
the former, we obtain
\begin{equation}
\Phi_A \equiv (\Phi, \Psi) = - k^{-2} \a \H^2 (3 \H \B + \b_A \E),
\quad \b_A = (1 - 4\b, 1).
 \label{eq:PP0}
\end{equation}

\subsection{Solution for a one-component medium}

Consider a universe filled with a one-component elastic medium
whose characteristics $p$, $\l$ and $\mu$ are all proportional to
$\r$. From equations (\ref{eq:Eeq}) we find
\begin{equation*}
\r \propto a^{-3w_+}, \quad a \propto \eta^{2u},
\end{equation*}
where $w_+ = 1 + w$ and $u = 1/(1 + 3w)$. (We suppose that $w >
-1/3$, otherwise we should write $a \propto (\mbox{sign\hskip
.2mm} u\ \eta)^{2u}$.) Note that since the compressional modulus
can be written as $K = dp/d\r\ \r_+$, it {\it must be}
proportional to $\r$, $K = w\r_+ = ww_+\r$. Thus, we must require,
besides that $w$ is constant, only that $\b$ is constant. If this
is the case, the functions appearing in (\ref{eq:dBdE}) are all
constant, except for the function $\cal H$ which is proportional
to $\eta^{-1}$,
\begin{equation*}
\a = \frac 32 w_+, \quad c_{S0}^2 = w, \quad c_{S\|}^2 = w + \frac
43 \b \equiv \tw, \quad \H = 2u \eta^{-1}.
\end{equation*}

After expressions for $\a$, $c_{S0}^2$, $c_{S\|}^2$ and $\cal H$
are inserted into equations (\ref{eq:dBdE}), they transform into
\begin{equation}
\B' = u (1 + 9w) \eta^{-1} \B + \tw \E, \quad \E' = - (k^2 + 18u^2
w_+ \eta^{-2}) \B - 3u w_+ \eta^{-1} \E.
 \label{eq:dBdEw}
\end{equation}
The two equations of first order for $\B$ and $\E$ can be combined
into one equation of second order for $\B$,
\begin{equation}
\B'' + 2\nu_B \eta^{-1} \B' + [q^2 - (2\nu_B - b)\eta^{-2}] \B =
0,
 \label{eq:ddB}
\end{equation}
where $q = \sqrt{\tw}k$, $\nu_B = u(1 - 3w)$ and $b = 24 u^2
w_+\b$. Furthermore, $\E$ can be expressed in terms of $\B$ and
$\B'$, and by using (\ref{eq:PP0}), $\Phi$ and $\Psi$ can be
expressed in terms of $\B$ and $\B'$, too. We obtain
\begin{equation}
\Phi_A  = - \b_A \S (q \eta)^{-2} [\B' - (1 - \s_A) \eta^{-1} \B],
 \label{eq:PP}
\end{equation}
where $\S = 6u^2 w_+$ and $\s_A = ((1 - 4\b)^{-1}, u)\ 8\b$.

Solution to equation (\ref{eq:ddB}) is
\begin{equation}
\B = z^{- \nu_-} (c_J J_n + c_Y Y_n),
 \label{eq:solB}
\end{equation}
where $z = q\eta$, $\nu_-$ and $n$ are defined in terms of $\nu =
\nu_B + 1/2 = 3u (1 - w)/2$ as $\nu_- = \nu - 1$ and $n =
\sqrt{\nu^2 - b}$, and $J_n$ and $Y_n$ are Bessel functions of
first and second kind of the argument $z$. Note that since $q =
c_{S\|}k$, the value $z = 1$ corresponds to the moment at which
the perturbation crosses the sound horizon (its reduced wavelength
becomes less than the radius of the horizon).

Knowing the function $\B$ we can determine the potentials $\Phi$
and $\Psi$. Denote $\nu_+ = \nu + 1$, $n_+ = n + 1$ and $m = \nu -
n$. By inserting (\ref{eq:solB}) into (\ref{eq:PP}) and using the
identities
\begin{equation*}
\frac {dJ_n}{dz} = - J_{n_+} + n z^{-1} J_n, \quad \frac
{dY_n}{dz} = - Y_{n_+} + n z^{-1} Y_n,
\end{equation*}
we obtain
\begin{equation}
\Phi_A = \b_A z^{-\nu_+} \{C_J [J_{n_+} + (m - \s_A) z^{-1}J_n] +
C_Y [Y_{n_+} + (m - \s_A) z^{-1}Y_n]\},
 \label{eq:solPP}
\end{equation}
where $C_J$ and $C_Y$ are defined in terms of $c_J$ and $c_Y$ as
$C_J = \S q c_J$ and $C_Y = \S q c_Y$. Another important quantity
is {\it density contrast} $\d = \overline{\d \r}/\r$. To determine
it, we need to know the function $\E$. By using the identities for
$dJ_n/dz$ and $dY_n/dz$ once again we find
\begin{equation}
\E = z^{-\nu_-} \{\hc_J [J_{n_+} + (m + \tau) z^{-1}J_n] + \hc_Y
[Y_{n_+} + (m + \tau) z^{-1}Y_n]\},
  \label{eq:E}
\end{equation}
where $\tau = 6uw$ and $\hc_J$ and $\hc_Y$ are defined in terms of
$c_J$ and $c_Y$ as $\hc_J = - \tw^{-1}q c_J$ and $\hc_Y = -
\tw^{-1}q c_Y$. The density contrast is obtained by inserting this
expression along with the expression for $\Psi \equiv \Phi_2$ into
\begin{equation}
\d = w_+ (3\Psi + \E).
 \label{eq:dw}
\end{equation}

An ideal fluid has $\b = 0$, hence $n = \nu$, $m = \s_A = 0$ and
\begin{equation}
\Phi = \Psi = z^{-\nu_+} (C_J J_{\nu_+} + C_Y Y_{\nu_+}).
 \label{eq:iPP}
\end{equation}
This agrees with the formula (7.58) in \cite{mukh} if we realize
that $z = \sqrt{w} k\eta$ for $\b = 0$ and $\nu_+ =$ $u (5 +
3w)/2$ for any $\b$.

Let us determine the asymptotics of the functions $\B$, $\Phi$,
$\Psi$ and $\E$ at $z \ll 1$ (in the first period after the
perturbation was formed, when its wavelength exceeded the size of
the horizon considerably). Denote the coefficients in the leading
terms in $J_n$ and $Y_n$ by $J$ and $Y$,
\begin{equation*}
J = \frac 1{2^n \G(n_+)}, \quad Y = -\frac 1 \pi 2^n \G(n),
\end{equation*}
and introduce one more parameter $M = \nu + n$. With these
notations we have
\begin{equation}
\B \doteq z (c_J J z^{-m} + c_Y Y z^{-M}).
 \label{eq:asB}
\end{equation}
Introduce, furthermore, the coefficients in the leading terms in
$J_{n_+}$ and $Y_{n_+}$, $J_+ = J/(2n_+)$ and $Y_+ = 2n Y$. Then
it holds
\begin{equation}
\Phi_A \doteq \b_A [C_J J_+ (P_A z^{-m} + Q_A z^{-2 - m}) + C_Y
Y_+ R_A z^{-2 - M}],
  \label{eq:asPP}
\end{equation}
where
\begin{equation*}
P_A = 1 - \frac 12 (m - \s_A), \quad Q_A = 2n_+ (m - \s_A), \quad
R_A = 1 + \frac 1{2n} (m - \s_A),
\end{equation*}
and
\begin{equation}
\E \doteq \hc_J J_+ Q z^{-m} + \hc_Y Y_+ R z^{- M},
  \label{eq:asE}
\end{equation}
where
\begin{equation*}
Q = 2n_+ (m + \tau), \quad R = 1 + \frac 1{2n} (m + \tau).
\end{equation*}
We have included the $P$-term into $\Phi_A$ although it is of
higher order in $z$ than the $Q$-term. The reason is that for
$|\b| \ll 1$, $P_A$ as well as $R_A \doteq 1$ while $Q_A = O(\b)$.
Thus, if $|\b|$ is small, the $P$-term can prevail over the
$Q$-term starting at some value of $z$ that is still small.

For the sake of completeness, note that the estimate of $Q_A$
refers to the case when $w$ is not close to zero. If it is, the
estimate is lower: for $|\b| \ll |w| \ll 1$ it holds $Q_A =
O(w\b)$ and for $|w| \ll |\b|$ it holds $Q_A = O(\b^2)$. This does
not affect the above argument, since it is using just the fact
that $Q_A$ is small for $|\b|$ small.

For an ideal fluid the asymptotics reduce to
\begin{equation}
\B \doteq z (c_J J + c_Y Y z^{-2\nu}),
  \label{eq:asBi}
\end{equation}
\begin{equation}
\Phi = \Psi \doteq C_J J_+ + C_Y Y_+ z^{-2\nu_+},
  \label{eq:asPPi}
\end{equation}
and
\begin{equation}
\E \doteq \hc_J J_+ Q_\i + \hc_Y Y_+ R_\i z^{- 2\nu},
  \label{eq:asEi}
\end{equation}
where $Q_{id}$ and $R_{id}$ are the values of $Q$ and $R$ for $\b
= 0$, $Q_\i = 2\nu_+ \tau$ and $R_\i = 1 + \tau/(2\nu)$.

\subsection{Switching the shear stress at a finite time}

Suppose the universe was originally filled with an ideal fluid
with the given value of $w$ and then, at some moment $\eta_s$, all
fluid instantaneously turned into a solid with the same $w$. Let
the transition be anisotropic, so that the solid was formed with
{\it Euclidean} internal geometry. Perturbations in such universe
are described by equations (\ref{eq:dBdE}), in which the parameter
$\b$ must be replaced by the function $\b \theta(\eta - \eta_s)$.
The equations imply that the functions $\B$ and $\E$ are both
continuous at $\eta_s$ and the derivative of $\E$ is continuous,
too, while the derivative of $\B$ has a jump coming from the jump
in $c_{S\|}^2 $. From equations (\ref{eq:Ps0}) and (\ref{eq:Ph0})
we can also see that the function $\Psi$ is continuous with a jump
in its derivative, while the function $\Phi$ has a jump itself.

The function $\B$ is given by equation (\ref{eq:solB}) both in the
ideal fluid and solid state era. However, in the latter era the
constants $c_J$ and $c_Y$, and even the variable $z$, are
different than in the former era. The constants change in order to
satisfy matching conditions and the variable changes because it
contains the parameter $q$ that switches from the value $q_0 =
\sqrt{w} k$ to the value $q = \sqrt{\tw} k$. Thus, if we write the
function $\B$ in the solid state era as in (\ref{eq:solB}), we
must write it in the ideal fluid era as
\begin{equation*}
\B_0 = z_0^{- \nu_-} (c_{J0} J_\nu + c_{Y0} Y_\nu),
\end{equation*}
where $z_0 = q_0 \eta$ and $J_\nu$ and $Y_\nu$ are Bessel
functions of first and second kind of the argument $z_0$. The
function $\E_0$, needed for the matching procedure, is obtained in
the same way from the ideal fluid version of (\ref{eq:E}).

The matching conditions read
\begin{equation}
\B_s = \B_{0s}, \quad \B'_s = \B'_{0s} + \frac 43 \b \E_{0s},
  \label{eq:match0}
\end{equation}
where the index $s$ indicates that the function is evaluated at
the moment $\eta_s$. The first condition states that $\B$ is
continuous at the moment $\eta_s$ and the second condition fixes
the jump in the derivative of $\B$ in accordance with the first
equation in (\ref{eq:dBdE}).

Suppose $z_{0s} = q_0 \eta_s \ll 1$; thus, the perturbation is by
assumption stretched far beyond the sound horizon at the moment
the shear stress switches on. Suppose, furthermore, that $z_s = q
\eta_s \ll 1$; thus, the perturbation stays stretched beyond the
sound horizon also during some period after the shear stress
switched on. (This is nontrivial in case $|w| \ll 1$ and $\b \sim
1$, since then $q_0 \ll q$.) The assumptions simplify the form of
the functions entering the matching conditions considerably. All
three are given by the asymptotic formulas valid at $z \ll 1$, the
functions $\B_0$ and $\E_0$ by equations (\ref{eq:asBi}) and
(\ref{eq:asEi}) and the function $\B$ by equation (\ref{eq:asB}).

The ratio of the second to the first term in the asymptotic
formulas for $\B_0$ and $\E_0$ varies with $z_0$ as $z_0^{-\nu}$.
Let $w$ be from the interval $(-1/3, 1)$. The parameter $\nu$ is
then positive and the ratio decays with time. Let us simplify the
theory even more by assuming that both terms were about the same
at the moment when the perturbation was formed, and that the shear
stress was switched long enough after that moment. Then we are
left with $\B_0$ and $\E_0$ containing the nondecaying term only,
\begin{equation}
\B_0 = c_{J0} J_0 z_0 \equiv C_0 z_0, \quad \E_0 = \hc_{J0} J_{0+}
Q_{\i, 0} = - 6u q_0 C_0.
 \label{eq:B0E0}
\end{equation}
After inserting these expressions into the combination of $\B_0'$
and $\E_0$ that appears on the right hand side of the second
matching condition, we find
\begin{equation*}
\B_0' + \frac 43 \b \E_0 = q_0 C_0 (1 - \s_2).
\end{equation*}
On the left hand side of the matching conditions we retain both
terms appearing in the asymptotic formula for $\B$, the term
proportional to $z^{1 - m}$ as well as the term proportional to
$z^{1 - M}$. As a result, we obtain
\begin{equation}
x + y = C, \quad (1 - m) x + (1 - M) y = C (1 - \s_2),
 \label{eq:match}
\end{equation}
where $C = q_0 C_0/q$ and $x$ and $y$ are defined in terms of
$c_J$ and $c_Y$ as $x = c_J J z_s^{-m}$ and $y = c_Y Y z_s^{-M}$.
The solution is
\begin{equation}
x = \frac C{2n} (M - \s_2), \quad y = - \frac C{2n} (m - \s_2).
 \label{eq:solxy}
\end{equation}

Instead of the constant $C$, it is more convenient to use
Newtonian potential of the perturbation in the ideal fluid era
$\Pz$. If we evaluate $\Pz$ under the same assumptions as $\B_0$
and $\E_0$, we obtain $\Pz = C_{J0} J_{0+}$, and since $C_{J0} =
\S q_0 c_{J0} = \S q_0 C_0/J_0 = \S qC/J_0$ and $J_{0+} =
J_0/(2\nu_+)$, we have
\begin{equation}
\Pz = \frac {\S q}{2\nu_+} C.
  \label{eq:P0}
\end{equation}
Using this relation together with the identities
\begin{equation*}
M - \s_2 = 2n R_2, \quad m - \s_2 = \frac 1{2n_+} Q_2,
\end{equation*}
we obtain
\begin{equation}
x = \frac {2\nu_+}{\S q} R_2 \Pz, \quad y = - \frac {\nu_+}{2n n_+
\S q} Q_2 \Pz.
 \label{eq:cJcY}
\end{equation}

For further reference, let us also express the density contrast in
the ideal fluid era $\d_0$ in terms of $\Pz$. It holds
\begin{equation*}
\E_0 = - \frac {12 u\nu_+}\S \Pz = - \frac {5 + 3w}{w_+} \Pz,
\end{equation*}
and by inserting this into the formula $\d_0 = w_+ (3\Phi_0 +
\E_0)$ we obtain, irrespective of the value of $w$,
\begin{equation}
\d_0 = - 2\Pz.
 \label{eq:d0P0}
\end{equation}

The asymptotics of $\Phi$, $\Psi$ and $\E$ can be found from
equations (\ref{eq:asPP}) and (\ref{eq:asE}) by computing $c_J J$
and $c_Y Y$ from $x$ and $y$ and using the formulas $C_J J_+ = \S
q c_J J/(2n_+)$, $C_Y Y_+ = 2n \S q c_Y Y$, and $\hc_J = - (\tw
\S)^{-1} C_J$, $\hc_Y = - (\tw \S)^{-1} C_Y$. If we also introduce
the rescaled time $\z = \eta/\eta_s = z/z_s$, we obtain
\begin{equation}
\Phi_A \doteq \P_A \z^{-m} + \Q_A z_s^{-2} \z^{-2 - m} + \R_A
z_s^{-2} \z^{-2 - M},
  \label{eq:asPPz}
\end{equation}
where
\begin{equation*}
\P_A = \b_A \frac{\nu_+}{n_+} P_A R_2 \Pz, \quad \Q_A = \b_A
\frac{\nu_+}{n_+} Q_A R_2 \Pz, \quad \R_A = - \b_A
\frac{\nu_+}{n_+} Q_2 R_A \Pz,
\end{equation*}
and
\begin{equation}
\E \doteq - \Q \z^{- m} - \R \z^{- M},
  \label{eq:asEz}
\end{equation}
where
\begin{equation*}
\Q = (\tw \S)^{-1} \frac{\nu_+}{n_+} Q R_2 \Pz, \quad \R = - (\tw
\S)^{-1} \frac{\nu_+}{n_+} Q_2 R \Pz.
\end{equation*}
Note that the coefficients $\Q_2$ and $\R_2$ in the expression for
$\Psi \equiv \Phi_2$ are the same except for their sign. This
guarantees that $\Psi$ does not have jump of order $z_s^{-2} \Pz$
at $\eta = \eta_s$.

In the formula for $\Phi_A$, the coefficients $\Q_A$ and $\R_A$
are multiplied by $z_s^{-2}$. Thus, at the moment $\eta_s$ the
$P$-term is suppressed with respect to the $Q$- and $R$-terms by
the factor $z_s^2$. On the other hand, if $|\b|$ is small, the
coefficients $\Q_A$ and $\R_A$ are of order $\b$, $\Q_A$ because
of the factor $Q_A$ appearing in the original formula and $\R_A$
because of the factor $Q_2$ coming from the expression for $c_J$.
Thus, at the moment $\eta_s$ the $P$-term is enhanced with respect
to the $Q$- and $R$-terms by the factor $\b^{-1}$. However, we
have solved the matching conditions only in the leading order in
$z_s$. The resulting theory is therefore applicable only if the
net effect is {\it suppression} of the $P$-term at the time
$\eta_s$; the term can eventually prevail, but only at times much
greater than $\eta_s$. This leads to the condition $|\b| \gg
z_s^2$. Since the estimate of $Q_A$ holds only for $w$ not too
close to zero, so does the constraint on $\b$. If we take into
account the behavior of $Q_A$ for $w$ close to zero, we arrive at
a stronger condition $|\b| \gg \min \{z_s^2/|w|, z_s\}$.

To justify our matching procedure, let us show that it is
consistent with the description of scalar perturbations in
Newtonian gauge. Equation with $\overline{\tau^{(1)}}$ on the
right hand side obtained in that gauge reads (see equation (7.40)
in \cite{mukh})
\begin{equation}
\Psi'' +  \H (2\Psi' + \Phi') + (2\H' + \H^2) \Psi - \frac 12
k^{2} (\Phi - \Psi) = - \frac 14\overline{\tau^{(1)}}.
  \label{eq:PPNewt}
\end{equation}
Denote the jump of the function at the moment $\eta_s$ by square
brackets. From equation (\ref{eq:PP0}) with $A = 1$ we find
\begin{equation*}
[\Phi] = 4k^{-2} \a_s \H_s^2 \b \E_s.
\end{equation*}
Equation (\ref{eq:PP0}) with $A = 2$ yields $[\Psi'] = - 3k^{-2}
\a_s \H_s^3 [\B']$, and if we use the formula $[\B'] = 4\b/3\
\E_s$, following from the first equation in (\ref{eq:dBdE}), we
have
\begin{equation}
[\Psi'] = - \H_s [\Phi].
 \label{eq:jump}
\end{equation}
The terms $\Psi''$ and $ \H \Phi'$ on the left hand side of
(\ref{eq:PPNewt}) both contain $\d$-function; however, the
identity we have obtained ensures that the $\d$-functions cancel
and only a jump-like discontinuity remains.


\section{Radiation-like solid}
 \label{sec:rad}

\subsection{Perturbations in a universe with radiation-like solid}

In standard cosmology the universe was dominated by the radiation
from the end of inflation almost up to recombination. Radiation is
ideal fluid with $w = 1/3$, therefore to make our problem more
realistic we must suppose that the value of $w$ in the ideal fluid
era was 1/3. Then we can use the previous theory without
modifications, if we require that the value of $w$ in the solid
state era was 1/3, too. This means that a portion of radiation has
been eventually converted into a solid with the same pressure to
energy density ratio. We will call such solid {\it
radiation-like}.

Evolution of perturbations in the presence of radiation-like solid
is given by the formulas derived in the previous section, with $w
= 1/3$ inserted everywhere. The constants entering the formulas
are $u = \nu = 1/2$, $\S = 2$, $\tau = 1$ and $b = 8\b$. If we
also rewrite $\b_A$, $\s_A$ and $\tw$ in terms of $b$ and insert
the value of $\nu$ into the definition of $n$, we obtain for the
remaining constants $\b_A = (1 - b/2, 1)$, $\s_A = ((1 -
b/2)^{-1}, 1/2) b$, $\tw = (1 + b/2)/3$ and $n = \sqrt{1/4 - b}$.

The parameter $\tw$ must be positive in order that longitudinal
sound waves are stable, and the parameter $n$ must be real in
order that the evolution of perturbations in a universe filled
with pure solid is smooth from the beginning. As a result, $b$
must be from the interval $(-2, 1/4)$. However, as we will see,
the theory agrees with observations only if $b$ is close to zero.
For such $b$ it holds $n \doteq 1/2 - b$, $m = 1/2 - n \doteq b$
and $M = 1/2 + n \doteq 1 - b$, and the constants in the
asymptotic formulas for the functions $\Phi$, $\Psi$ and $\E$,
evaluated in the leading order in $b$, are
\begin{equation*}
(P_1, Q_1, R_1) \doteq \Bl 1, \frac 32 b^2, 1\Br, \quad (P_2, Q_2,
R_2) \doteq \Bl 1, \frac 32 b, 1\Br, \quad (Q, R) \doteq (3, 2).
\end{equation*}
While $Q_2$ is of first order in $b$ as expected, $Q_1$ turns out
to be of {\it second} order. The value $w = 1/3$ is special in
this respect. (The only other value for which $Q_1$ is of second
order is $w = 0$, but $Q_2$ is for that $w$ of second order, too.)
When arguing that the constraint $|b| \gg z_s^2$ must be observed
in order that the expressions for $(P_A, Q_A, R_A)$ are valid, we
have assumed that both $Q_A$ are of first order in $b$. The fact
that $Q_1$ is of second order does not lead to strengthening of
this constraint; expressions for $(P_1, Q_1, R_1)$ can be used
also for $|b| \lesssim z_s$ because the $P$-term in $\Phi$, while
not suppressed with respect to the $Q$-term at the moment
$\eta_s$, is still suppressed with respect to the $R$-term. After
inserting these expressions into the formulas for $(\P_A, \Q_A,
\R_A)$ and $(\Q, \R)$ and using the approximate equalities $\b_A
\nu_+/n_+ \doteq (1, 1)$ and $(\tw \S)^{-1} \nu_+/n_+ \doteq 3/2$,
we obtain
\begin{equation*}
(\P_1, \Q_1, \R_1) \doteq \Bl 1, \frac 32 b^2, - \frac 32 b\Br
\Pz, \quad (\P_2, \Q_2, \R_2) \doteq \Bl 1, \frac 32 b, - \frac 32
b\Br \Pz, \quad (\Q, \R) \doteq \frac 92 (1, - b) \Pz.
\end{equation*}
As a result, approximate expressions for the functions $\Phi$,
$\Psi$ and $\E$ in the regime in which the perturbation is
stretched far beyond the sound horizon are
\begin{equation}
\Phi \doteq \BL \z^{-b} + \frac 32  b z_s^{-2} (b\z^{-2 - b} -
\z^{-3 + b})\BR \Pz, \quad \Psi \doteq \BL \z^{-b} + \frac 32  b
z_s^{-2} (\z^{-2 - b} - \z^{-3 + b})\BR \Pz,
  \label{eq:appp}
\end{equation}
and
\begin{equation}
\E \doteq - \frac 92 (\z^{- b} - b \z^{-1 + b}) \Pz.
  \label{eq:ape}
\end{equation}
Knowing the functions $\Psi$ and $\E$, we can compute the density
contrast as $\d = 4(\Psi + \E/3)$.

The dependence of the functions $\tP = \Phi/\Pz$, $\tPs =
\Psi/\Pz$ and $\td = \d/\Pz$ on time is shown in fig. 1. The
rescaled density contrast is multiplied by $-1/2$ in order to
normalize it to 1 in the ideal fluid era
\begin{figure}[ht]
\centerline{\includegraphics[width=0.9\textwidth]{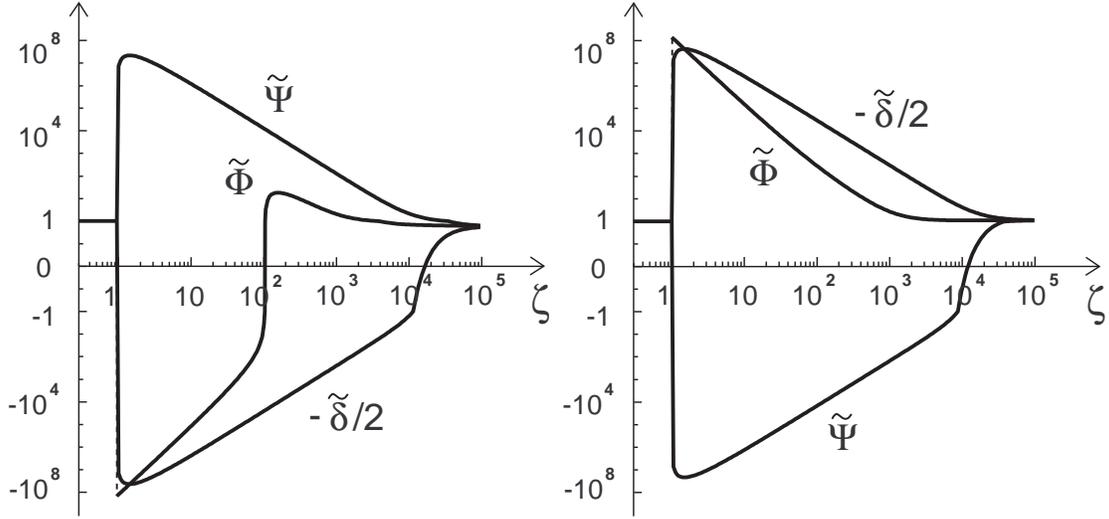}}
\centerline{\parbox {12.2cm} {\caption{\small Behavior of
gravitational potentials and density contrast for positive (left)
and negative (right) shear stress}}}
  \label{fig:PPE}
\end{figure}
(see equation (\ref{eq:d0P0})). The value of the dimensionless
shear modulus is $b = 0.01$ in the left panel and $b = -0.01$ in
the right panel, and the value of the variable $z$ at the moment
when the shear stress is switched on is $z_s = 10^{-5}$ in both
panels. The scale is linear in the central band and logarithmic
outside of it. The variable $\z$ is bounded from above by the
value $z_s^{-1} = 10^5$, at which $z = 1$ and the perturbation
crosses the sound horizon. Up to that point, we have evaluated the
functions $\tP$, $\tPs$ and $\tE$ from the asymptotic formulas
(\ref{eq:appp}) and (\ref{eq:ape}) with suppressed factor $\Pz$.
This is an extrapolation, since the formulas are applicable only
at $z \ll 1$. At $z \sim 1$ the functions begin to oscillate;
thus, the curves computed from exact formulas bend down at the
right edge of the panels instead of being approximately
horizontal.

The function $\tP$ jumps from 1 to $- 3b/2\ z_s^{-2} = \mp 1.5
\tim 10^8$ at $\z = 1$. (All expressions in this paragraph are
approximate, with the leading term cited only.) The function
$\tPs$ is continuous, however, it rises abruptly with rising $\z$
in case $b > 0$ and falls down abruptly with rising $\z$ in case
$b < 0$, so that it assumes a comparable value with opposite sign
at a nearby $\z$. Its maximum/minimum is reached at $\z = 3/2$ and
equals $2b/9\ z_s^{-2} = \pm 2.2 \tim 10^7$. As $\z$ increases,
$\tPs$ decreases as $\z^{-2 - b}$ in case $b > 0$ and increases as
$- \z^{-2 - b}$ in case $b < 0$, and then, after $\z$ reaches the
value $|b|^{1/2} z_s^{-1} = 10^4$, it relaxes to the function
$\z^{-b}$, which is approximately constant and equal to 1 in the
interval of $\z$ under consideration. The function $- \td/2$
coincides approximately with $- \tPs$ as long as $\z$ stays less
than $|b|^{1/2} z_s^{-1}$. Thus, it reaches the value $\mp 4.4
\tim 10^7$ at $\z = 3/2$ and then it increases as $- \z^{-2 - b}$
in case $b > 0$ and decreases as $\z^{-2 - b}$ in case $b < 0$.
After $\z$ rises above $|b|^{1/2} z_s^{-1}$, it relaxes to the
function $\z^{-b}$ just as $\tPs$ does. We can summarize this
behavior by saying that $\td$ switches from $2\tPs$ in the first
regime to $-2\tPs$ in the second regime. Finally, the function
$\tP$ operates in three regimes. For $\z \lesssim |b|^{-1} = 100$
it increases as $- \z^{-3 + b}$ in case $b > 0$ and decreases as
$\z^{-3 + b}$ in case $b < 0$; for $|b|^{-1} \lesssim \z \lesssim
|b| z_s^{-1} = 1000$ it decreases as $\z^{-2 - b}$ in both cases;
and for $\z \gtrsim |b| z_s^{-1}$ it relaxes to $\z^{-b}$. In case
$b > 0$, it reaches maximum when passing from the first regime to
the second one. The maximum occurs at $\z = 3/(2b) = 150$ and the
value of $\tP$ is $2 b^4/9\ z_s^{-2} = 22$.

\subsection{Size of perturbations}

Perturbations must be small in order that the linearized theory
describing them is applicable. However, the size of the
perturbations depends on gauge. By changing it, one can surely
turn a small perturbation into a large one. Therefore the
smallness of perturbations must be defined as an existence
property: the perturbation is small if there {\it exists} gauge in
which it is small.

To estimate the size of perturbations we need to know the value of
$\Pz$. Consider the perturbed universe at the moment of
recombination $\eta_{re}$. We can describe the behavior of
perturbations up to that moment {\it approximately} by the
asymptotic formulas (\ref{eq:appp}) and (\ref{eq:ape}). The
description is approximate since we are ignoring the fact that the
universe becomes matter-dominated in the last period before
$\eta_{re}$. Because of that, the parameter $w$ does not stay
constant; it falls down from 1/3 to approximately 1/12. Suppose
the perturbation crosses the sound horizon at recombination; in
other words, suppose the wave number $k$ is such that the moment
of horizon crossing $\eta_h = q^{-1} = (\tw k)^{-1}$ equals
$\eta_{re}$. As seen from fig. 1, this yields values of
$\Phi_{re}$, $\Psi_{re}$ and $-\d_{re}/2$ close to $\Phi_0$
(because the values of $\tP$, $\tPs$ and $-\td/2$ at $\z =
z_s^{-1}$ are close to 1). On the other hand, from the magnitude
of CMB anisotropies we know that the quantities $\Phi_{re}$,
$\Psi_{re}$ and $\d_{re}$ are of order 10$^{-5}$ for
long-wavelength perturbation. Thus, the value of $\Pz$ for the
perturbations under consideration must be of order 10$^{-5}$, too.
Inflation yields flat initial spectrum, so that if we accept
inflation as the mechanism by which the perturbations were
created, $\Pz$ must be of order 10$^{-5}$ for {\it all}
perturbations.

The maximum the functions $\Phi$, $\Psi$ and $\d$ reach in
absolute value is of order $|b| z_s^{-2} \Pz$. (We suppose $\Pz >
0$.) As a result, if we describe the perturbed universe by these
functions, the requirement that the perturbation is small leads to
the constraint on the dimensionless shear modulus $|b| \ll z_s^2
\Pz^{-1} \approx 10^5 z_s^2$. To get an idea of how strong the
constraint is, consider perturbations crossing the sound horizon
at recombination. By definition, the parameter $z_s$ equals $q
\eta_s$, or $\eta_s/\eta_h$, hence its value for such
perturbations is $z_s^\0 = \eta_s/\eta_{re}$. The time dependence
of the scale parameter in a universe filled with radiation is $a
\propto \eta$, so that $z_s^\0 = a_s/a_{re}$; or, if we denote the
temperature of cosmic medium by $T$, $z_s^\0 = T_{re}/T_s$. Thus,
$z_s^\0$ is the ratio of two energy scales, the scale of
recombination and the scale of shear switch. The value $z_s =
10^{-5}$ we have used in our illustrative computation, if
identified with $z_s^\0$, corresponds with the scale of shear
switch 0.1~MeV. In other words, $z_s^\0$ assumes this value if the
radiation-like solid has been formed not earlier than at the time
of nucleosynthesis. The resulting constraint on the shear stress
is $|b| \ll 10^{-5}$. If the solid appears at an earlier stage,
the constraint becomes stronger. In particular, formation of the
solid on the GUT scale leads to $z_s^\0 = 10^{-23}$ and $|b| \ll
10^{-41}$.

The previous analysis can be easily extended to long-wavelength
perturbations. If we denote the wave number of perturbations
crossing the sound horizon at recombination by $k^\0$, the
perturbations with the longest wavelength that can be observed in
CMB have $k \doteq 0.01 k^\0$. Thus, their $z_s$ is by two orders
of magnitude less than $z_s^\0$, and to keep them small, we must
restrict the value of $|b|$ by a number that is by four orders of
magnitude smaller than the numbers cited above. We can see that to
guarantee that the functions $\Phi$, $\Psi$ and $\d$ are close to
zero, the universe must be filled with matter that is practically
indistinguishable from an ideal fluid.

The functions $\Phi$, $\Psi$ and $\d$ have been defined in a gauge
invariant way, but their invariance is {\it computational}, not
{\it conceptual}. They can be calculated in any gauge by using the
formulas (\ref{eq:db}), but they refer to a {\it particular}
gauge, namely to Newtonian gauge defined by the conditions $h_{0i}
= E = 0$. Thus, the constraints we have established by requiring
that $\Phi$, $\Psi$ and $\d$ are small in absolute value are in
fact gauge dependent. To find out whether they cannot be relaxed,
let us look at the constraints in the proper-time comoving gauge,
defined by the conditions $\phi = \bm \xi = 0$.

The scalar part of the perturbation to the metric in the
proper-time comoving gauge is given by the three functions $kB$
(since $h^{(S)}_{0i} = B_{,i} = ikB$), $\psi$ and $\E = -k^2 E$
(since $h^{(S)}_{ij}$ is the sum of $h^{(S1)}_{ij} = 2\psi
\d_{ij}$ and $h^{(S2)}_{ij} = 2E_{,ij} = -2k_i k_j E$).
Perturbation to the matter density equals $4(\psi + \E/3)$ and
does not need to be considered separately. For the function $\E$
we have expression (\ref{eq:asEz}), which for $w = 1/3$ and $|b|
\ll 1$ reduces to (\ref{eq:ape}). The function $k\B$ is given by
equation (\ref{eq:asB}), which transforms after inserting for
$c_J$ and $c_Y$ into
\begin{equation}
k\B \doteq z(\Q_b \z^{- m} + \R_b \z^{- M}),
  \label{eq:asBz}
\end{equation}
where
\begin{equation*}
\Q_b = (\sqrt{\tw} \S)^{-1} 2\nu_+ R_2 \Pz, \quad \R_b = -
(\sqrt{\tw} \S)^{-1} \frac{\nu_+}{2n n_+} Q_2 \Pz,
\end{equation*}
For $w = 1/3$ and $|b| \ll 1$ this reduces to
\begin{equation}
k\B \doteq \frac {3\sqrt{3}}2 z\Bl \z^{- b} - \frac 12 b \z^{-1 +
b}\Br \Pz.
  \label{eq:apb}
\end{equation}
The function $\chi$ appearing in the expression for $kB$ satisfies
an equation coming from the longitudinal part of the equation
$2G_{0i} = T_{0i}$. It can be obtained from the equation for $y$
in \cite{pol} by putting $y = a\chi$, and is of the form
\begin{equation*}
\chi' = - \H (\chi + \a \B).
\end{equation*}
The solution is
\begin{equation}
\chi = - a^{-1} \int \a \B\ da,
 \label{eq:chi0}
\end{equation}
and if we insert here $\a = 2$ and $a \propto \eta$ and use the
approximate expression (\ref{eq:apb}) for $k\B$, we obtain
\begin{equation}
k \chi \doteq - \frac {3\sqrt{3}}2 z \BL \Bl 1 + \frac b2 \Br
\z^{- b} - b \z^{-1 + b})\BR \Pz.
 \label{eq:chi}
\end{equation}
We have skipped the term with integration constant since it can be
always removed by an appropriate choice of the start of time
counting. On the other hand, we have included correction of order
$b$ into the term in square brackets proportional to $\z^{- b}$,
in order that we are able to calculate the leading term
proportional to $\z^{- b}$ in the function $kB = k(\B + \chi)$. Of
course, we do not need this correction when computing the function
$\psi = - \eta^{-1} \chi = - \sqrt{\tw} z^{-1} k\chi$. After
inserting into the definitions of $kB$ and $\psi$ from equations
(\ref{eq:apb}) and (\ref{eq:chi}), we find
\begin{equation}
kB \doteq - \frac {3\sqrt{3}}4 b z(\z^{- b} - \z^{-1 + b}) \Pz,
\quad \psi \doteq \frac 32 (\z^{- b} - b \z^{-1 + b}) \Pz.
 \label{eq:apbp}
\end{equation}

We are interested in the behavior of the functions $kB$, $\psi$
and $\E$ for $\z$ ranging from 1 (the time the solid appeared) to
$(z_s^\0)^{-1}$ (the time of recombination). The formulas
(\ref{eq:ape}) and (\ref{eq:apbp}) hold, with greater or less
accuracy, on the whole interval of $\z$ if $k \le k^\0$, and up to
$\z = z_s^{-1} = (k/k^\0)^{-1} \tim$ the maximum $\z$ if $k
> k^\0$. Outside that interval all three functions oscillate,
$kB$ and $\E$ with constant amplitude and $\psi$ with falling
amplitude. (The two terms in $k\B$ and the leading two terms in
$\E$ are of the form $z^{1/2} \tim$ Bessel function, and Bessel
functions oscillate with the amplitude $\sim z^{-1/2}$; thus,
$k\B$ and $\E$ oscillate with constant amplitude, $\chi$
oscillates with amplitude proportional to $z^{-1}$ and $\psi$
oscillates with amplitude proportional to $z^{-2}$.) Clearly, if
we want to estimate the functions $kB$, $\psi$ and $\E$ in
absolute value from above, for $k > k^\0$ we can use the same
formulas as for $k \le k^\0$, only on a smaller interval. We will
restrict ourselves to perturbations with $k = k^\0$; however, it
can be easily checked by using the effective domain of $kB$,
$\psi$ and $\E$ that the results stay the same after one extends
the analysis to perturbations with arbitrary $k$.

The functions $\z^{- b}$ and $\z^{-1 + b}$ both equal 1 at $\z =
1$, and since $|b| \ll 1$, the second function always decreases
while the first function either decreases at a slower rate or
increases. Thus, the first term in $kB$ dominates the second term
except for values of $\z$ comparable with 1, and the first term in
$\psi$ and $\E$ dominates the second term for all $\z$. The order
of magnitude of $kB$ is given by its first term even for $\z$ of
order 1, except for a small interval close to 1 in which $kB$ is
close to zero. As a result, $kB$ is in absolute value almost
everywhere of order $|b| z \z^{- b} \Pz$ and $\psi$ and $\E$ are
in absolute value everywhere of order $\z^{- b} \Pz$. If $b > 0$,
the maximum of the function $z \z^{- b} = z_s^\0 \z^{1 - b}$ for
$k = k^\0$ is $(z_s^\0)^b$, which is less than 1, and the maximum
of the function $\z^{- b}$ is 1; if $b < 0$, the maximum of both
functions is $(z_s^\0)^b$, which is greater than 1. We arrive at
the conclusion that for $b > 0$ the requirement of smallness of
perturbations does not give any constraint on $b$, while for $b <
0$ it leads to the constraint $(z_s^\0)^b \ll \Pz^{-1} \approx
10^5$, or $b \gtrsim 5/\log_{10} (z_s^\0)$. If we extrapolate the
theory to finite values of $b$, we find that $b$ must be slightly
greater than $-1$ if the solid appeared at the scale of
nucleosynthesis, and slightly greater than $-0.22$ if it appeared
at the GUT scale. These conditions are {\it substantially} weaker
than the ones we have obtained in Newtonian gauge.

The analysis of the behavior of the functions $kB$, $\psi$ and
$\E$ shows that if the functions $\Phi$, $\Psi$ and $\d$ in a
universe with solid component become in absolute value comparable
with 1, or even much greater than 1, that does not mean that the
theory has collapsed. Such behavior means simply that Newtonian
gauge is not appropriate for the description of perturbations at
the given stage of the evolution of the universe. Of course, it
can be still used during the stages when the functions $\Phi$,
$\Psi$ and $\d$ are small in absolute value (before the solid
component was formed as well as in the last period before the
perturbation entered the horizon).

Newtonian gauge is usually viewed as closest to intuition, since
the theory resembles Newton's theory of gravitation in it.
However, this does not hold for superhorizon perturbations, which
are substantially non-Newtonian since their properties do
crucially depend on the finiteness of the speed of propagation of
electromagnetic as well as gravitational interaction. As a result,
even if there exists an intuitive explanation of the failure of
Newtonian gauge in the first period after the shear stress has
been switched, it is by no means self-evident. As it looks without
further analysis, the failure is just a consequence of the
peculiar character of coordinate transformations in which the new
coordinates $x^\mu$ depend weakly on the old coordinates $\bf x$
($\d x^\mu \propto e^{i {\bf k} \cdot {\bf x}}$ with $k\eta \ll
1$).

Even if the perturbations are small, the theory must be ruled out
if it contradicts observations. This is where the restriction on
the values of $b$ used throughout this section comes from.
Consider perturbations corresponding to the large-scale part of
CMB anisotropies, $0.01 k^\0 < k < k^\0$. Observations suggest
that the quantities $\Phi_{re}$, $\Psi_{re}$ and $\d_{re}$ for
such values of $k$ are approximately constant. (The exact
formulation of that is that the product of Fourier components of
$\Phi_{re}$, $\Psi_{re}$ and $\d_{re}$ as functions of $\bf x$,
averaged over the ensemble of universes, is approximately
proportional to $k^3 \d ({\bf k} - {\bf k}')$.) The quantities
$\Phi_{re}$, $\Psi_{re}$ and $\d_{re}$ are the values of the
functions $\Phi$, $\Psi$ and $\d$ at the moment $\z_{re} =
(z_s^\0)^{-1} = k/k^\0\ z_s^{-1}$. Thus, we require that the three
functions are approximately constant throughout the interval $0.01
z_s^{-1} < \z < z_s^{-1}$.

If $|b| \ll 1$, the functions $\Phi$, $\Psi$ and $\d$ are varying
slowly as long as they are dominated by the term proportional to
$\z^{- b}$. The first function passes to this regime at $\z
\approx |b| z_s^{-1}$ and the other two functions pass to this
regime at $\z \approx |b|^{1/2} z_s^{-1}$. This is confirmed by a
more detailed analysis taking into account the presence of matter
in cosmic medium. (In fact, even at $\z \ll \z_{re}$, when the
effect of matter is small, the exact expression for $\Phi$ looks
different than in (\ref{eq:appp}). It has the factor $b$ in front
of $\z^{-2 - b}$ replaced by $b + \z/(2\z_{re})$. However, the
correction starts to be important not earlier than at $\z \approx
|b| \z_{re} = |b| z_s^{-1}$, which is just the moment when the
term proportional to $\z^{-2 - b}$ starts to be dominated by the
term proportional to $\z^{- b}$.) Since we require that all {\it
three} functions are varying slowly, the relevant interval of $\z$
is the less of the two, $|b|^{1/2} z_s^{-1} \lesssim \z <
z_s^{-1}$. In order that this is contained in the interval $0.01
z_s^{-1} < \z < z_s^{-1}$, the parameter $b$ must satisfy $|b|
\lesssim 10^{-4}$. A natural description of shear stress is with
the help of the parameter $\xi = w_+ \b$, equal to $b/6$ in case
$w = 1/3$. The observational constraint on this parameter is
$|\xi| \lesssim 10^{-5}$.

\section{Stiff solid}
 \label{sec:stiff}

\subsection{Expansion of a universe with stiff solid}

Suppose a solid with $w > 1/3$ appears in a universe filled with
radiation. Density of matter decreases with the increasing scale
parameter as $a^{-3w_+}$, the faster the greater the value of $w$.
Thus, if the solid acquires a substantial part of the energy of
radiation at the moment it is formed, it will dominate the
evolution of the universe for a limited period until the radiation
takes over again. Let us determine how the dynamics of such
universe looks like.

Denote the part of the energy of radiation that transfers to the
solid by $1 - \e$. In the period with pure radiation ($\eta <
\eta_s$) the mass density is $\r = \r_s (a_s/a)^4$, so that the
first equation in (\ref{eq:Eeq}) yields
\begin{equation}
a = C \eta, \quad C = \Bl\frac 16 \r_s a_s^4\Br^{1/2}.
 \label{eq:arad}
\end{equation}
In the period with a mix of radiation and solid ($\eta > \eta_s$)
the mass density is
\begin{equation*}
\r = \e \r_s (a_s/a)^4  + (1 - \e) \r_s (a_s/a)^{3w_+} = \r_s
(a_s/a)^4 [\e + (1 - \e)(a_s/a)^\D],
\end{equation*}
where $\D = 3w_+ - 4$. As a result, the first equation in
(\ref{eq:Eeq}) transforms into
\begin{equation}
a' = C [\e + (1 - \e)(a_s/a)^\D]^{1/2}.
 \label{eq:aseq}
\end{equation}
In the interval of $w$ we are interested in the parameter $\D$ is
positive, therefore the second term eventually becomes less than
the first term even if $\e \ll 1$.

Equation (\ref{eq:aseq}) solves analytically for $w = 2/3$ and $w
= 1$, when $\D = 1$ and $\D = 2$. Note that for $w = 1$ it holds
$\nu = 0$ and $n = \sqrt{-b}$; thus, in a universe filled with
pure solid with $w = 1$, evolution of perturbations is smooth from
the moment the solid was formed on only if the shear stress is
negative. The solution is, for $w = 2/3$,
\begin{equation}
a = \frac {1 - \e}{2\e} a_s (\cosh \psi - 1), \quad C\te = \frac
{1 - \e}{2\e \sqrt{\e}} a_s (\sinh \psi - \psi),
 \label{eq:as23}
\end{equation}
and for $w = 1$,
\begin{equation}
a = (\e C^2 \te^2 + 2a_s \sqrt{1 - \e} C \te)^{1/2}.
 \label{eq:as1}
\end{equation}
In both formulas there appears {\it shifted time} $\te = \eta -
\eta_*$, with $\eta_*$ defined in such a way that the resulting
function $a(\eta)$ matches the function (\ref{eq:arad}) at $\eta =
\eta_s$. The shift is given for $w = 2/3$ by
\begin{equation}
\eta_* = \BL 1 - \frac 1\e \Bl 1 - \frac {1 - \e}{2\sqrt{\e}} \log
\frac {1 + \sqrt{\e}}{1 - \sqrt{\e}} \Br\BR \eta_s,
 \label{eq:shift23}
\end{equation}
and for $w = 1$ by
\begin{equation}
\eta_* = \Bl 1 - \frac {1 - \sqrt{1 - \e}}\e\Br \eta_s.
 \label{eq:shift1}
\end{equation}

In what follows we will use, instead of exact solutions for
special $w$'s and any $\e$, approximate solution for any $w$ and
$\e \ll 1$. Suppose less than one half of the total energy remains
stored in radiation at the moment of radiation-to-solid transition
($\e < 1/2$). The subsequent expansion of the universe can be
divided into two eras, solid dominated and radiation dominated,
separated by the time $\eta_{rad}$ at which the mass densities of
the solid and radiation are the same. The value of $\eta_{rad}$ is
given by
\begin{equation}
a_{rad} = a_s (\e^{-1} - 1)^{1/\D}.
 \label{eq:aeq}
\end{equation}
Suppose now that the post-transitional share of energy stored in
radiation is small ($\e \ll 1$). The universe then expands by a
large factor between the times $\eta_s$ and $\eta_{rad}$,
\begin{equation*}
a_{rad} \doteq a_s \e^{-1/\D} \gg a_s,
\end{equation*}
and we can be describe it in a good approximation as if it was
filled first with pure solid and then with pure radiation. Thus,
we replace equation (\ref{eq:aseq}) by
\begin{equation}
a' \doteq \bigg\{ \mbox{\hskip -2mm}
  \left. \begin{array} {l}
  C (a_s/a)^{\D/2} \mbox{ for } \eta < \eta_{rad}\\
  \sqrt{\e}C \mbox{ for } \eta > \eta_{rad}\\
  \end{array}\mbox{\hskip -1mm}. \right.
 \label{eq:aseqap}
\end{equation}
The solution is
\begin{equation}
a \doteq \bigg\{ \mbox{\hskip -2mm}
  \left. \begin{array} {l}
  \big[(\D/2 + 1) a_s^{\D/2} C\te\big]^{\frac 1{\D/2 + 1}} \mbox{ for }
  \eta < \eta_{rad}\\
  \sqrt{\e} C \tte \mbox{ for } \eta > \eta_{rad}\\
  \end{array}\mbox{\hskip -1mm}, \right.
 \label{eq:asap}
\end{equation}
where $\te$ and $\tte$ are shifted time variables, $\te = \eta -
\eta_*$ and $\tte = \te - \eta_{**}$. From the approximate
expression for $a_{rad}$ we obtain
\begin{equation}
\te_{rad} = \frac 1{\D/2 + 1} \e^{- \frac {\D/2 + 1}\D} \eta_s,
 \label{eq:eeq}
\end{equation}
and by matching the solutions at $\eta_s$ and $\eta_{rad}$ we find
\begin{equation}
\eta_* = \frac {\D/2}{\D/2 + 1} \eta_s, \quad \eta_{**} = - \frac
\D2 \te_{rad},
 \label{eq:eestar}
\end{equation}
As a quick test of the exact solutions cited above we can check
that expressions (\ref{eq:shift23}) and (\ref{eq:shift1}) for
$\eta_*$ reduce to the first expression in (\ref{eq:eestar}) in
the limit $\e \ll 1$.

The two equations in (\ref{eq:eestar}) can be rewritten as
\begin{equation*}
\frac {\te_s}{\eta_s} = \frac 1{\D/2 + 1} = \frac u{u_0}, \quad
\frac {\tte_{rad}}{\te_{rad}} = \frac \D2 + 1 = \frac {u_0}u,
\end{equation*}
where $u_0$ is the value of $u$ in the radiation era. (We have
used that $u = 1/(\D + 2)$ and $u_0 = 1/2$.) Expressions for the
ratios $\te_s/\eta_s$ and $\tte_{rad}/\te_{rad}$ in terms of the
ratio $u/u_0$ stay valid also after we replace radiation by an
ideal fluid with arbitrary pressure-to-radiation ratio $w_0$. To
demonstrate that, let us derive them from the condition of
continuity of Hubble parameter. If in the given period of time the
universe is filled with matter with the given value of $w$, its
scale parameter depends on a suitably shifted time $\te$ as $a
\propto \te^{2u}$. Thus, its Hubble parameter is $\H = 2u
\te^{-1}$ and the requirement that $\H$ is continuous at the
moment when $w$ changes from $w_I$ to $w_\I$ is equivalent to
$\te_\I/\te_I = u_\I/u_I$.

\subsection{Transitions with jump in $w$}

Suppose the functions $w_\eta$ and $\b_\eta$ change at the given
moment $\eta_{tr}$ (``transition time'') from $(w_I, \b_I)$ to
$(w_\I, \b_\I) = (w_I + \D w, \b_I + \D \b)$. (We have attached
the index $\eta$ to the symbols $w$ and $\b$ in order to
distinguish the functions denoted by them from the values these
functions assume in a particular era.) Rewrite the first equation
in (\ref{eq:dBdE}) as
\begin{equation}
\B' = c_{S0}^2 (3\H \B + \E) + \Bl\frac 32 w_{\eta +} - 1\Br\H \B
+ \frac 43 \b_\eta \E,
 \label{eq:Brew}
\end{equation}
where
\begin{equation}
c_{S0}^2 = \frac {dp}{d\r} = w_\eta + \r \frac {dw_\eta}{d\r}.
 \label{eq:cs02}
\end{equation}
Because of the jump in $w_\eta$ there appears $\d$-function in
$c_{S0}^2$, and to account for it, we must assume that $\B$ has a
jump, too. However, on the right hand side of equation
(\ref{eq:Brew}) we then obtain an expression of the form
``$\theta$-function$\tim \d$-function''; and if we rewrite $\B'$
as
\begin{equation*}
\B' = \frac {d\B}{d\r} \r' = - 3\H \r w_{\eta +} \frac {d\B}{d\r},
\end{equation*}
on the left hand side there appears another such expression. To
give meaning to the equation we must suppose that $w_\eta$ changes
from $w_I$ to $w_\I$ within an interval of the length $\D \r \ll
\r_{tr}$, and send $\D \r$ to zero in the end. If we retain just
the leading terms in equation (\ref{eq:Brew}) in the interval
under consideration, we obtain
\begin{equation}
w_{\eta +} \frac {d\B}{d\r} = -\Bl \B + \frac
{\E_{tr}}{3\H_{tr}}\Br \frac {dw_\eta}{d\r},
 \label{eq:Blead}
\end{equation}
where we have used the fact that, as seen from the second equation
in (\ref{eq:dBdE}), the function $\E$ is continuous at $\eta =
\eta_{tr}$. The solution is
\begin{equation*}
\B + \frac {\E_{tr}}{3\H_{tr}} = \frac \C{w_{\eta +}}.
\end{equation*}
To compute the jump in $\B$, we express $\B_I$ and $\B_\I$ in
terms of $w_{I +}$ and $w_{\I +}$, compute the difference $\B_\I -
\B_I$ and use the expression for $\B_I$ to exclude $\C$. In this
way we find
\begin{equation}
[\B] = - \frac {\D w}{w_{\I +}} \Bl \B_I + \frac
{\E_{tr}}{3\H_{tr}} \Br.
 \label{eq:jB}
\end{equation}
Note that the same formula is obtained if we assume that the
functions with jump are equal to the mean of their limits from the
left and from the right at the point where the jump occurs.

To justify the expression for $[\B]$ we can compute the jump in
$\Psi$,
\begin{equation*}
[\Psi] =  - \frac 32 k^{-2} \H_{tr}^2 (3 \H_{tr} [w_{\eta+} \B] +
\D w \E_{tr}).
\end{equation*}
If we write $[w_{\eta+} \B] = w_{\I+} [\B] + \D w \B_I$ and insert
for $[\B]$, we immediately see that $[\Psi]$ vanishes. This must
be so because a jump in $\Psi$ would produce a derivative of
$\d$-function in equation (\ref{eq:PPNewt}), and no such
expression with opposite sign appears in the other terms present
there.

The jump in $\B'$ can be found from equation (\ref{eq:Brew}) by
computing the jump of the right hand side, with no need for the
limiting procedure we have used when determining the jump in $\B$.
The result is
\begin{equation}
[\B'] = 4 \frac {\D w}{w_{\I+}} \H_{tr} \B_{tr} + \Bl \frac {5 -
3w_\I}{6w_{\I+}} \D w + \frac 43 \D \b \Br \E_{tr}.
 \label{eq:jdB}
\end{equation}

\subsection{Perturbations in a universe with stiff solid}

We are interested in perturbations in a universe in which the
parameters $w$ and $\b$ assume values $(w_0, 0)$ before $\eta_s$,
$(w, \b)$ between $\eta_s$ and $\eta_{rad}$, and $(w_0, 0)$ after
$\eta_{rad}$. (For most of this subsection we will leave $w_0$
free, only at the end we will put $w_0 = 1/3$.) Denote the
functions describing the perturbation before $\eta_s$ by the index
0, between $\eta_s$ and $\eta_{rad}$ by the index $s$, and after
$\eta_{rad}$ by the index 1. If only the nondecaying part of
perturbation survives before the moment when the solid appears,
$\B_0$ and $\E_0$ are given by expressions (\ref{eq:B0E0}) with
$u$ replaced by $u_0$ and $q_0$ defined as $\sqrt{w_0}k$. If,
furthermore, the perturbation is stretched far beyond the horizon
all the time, $\B_s$ and $\E_s$ are given by expressions
(\ref{eq:asB}) and (\ref{eq:asE}) with $z$ replaced by $\tz = q
\te$, and $\B_1$ is given by expression (\ref{eq:asBi}) with $c_J$
and $c_Y$ replaced by $c_{J1}$ and $c_{Y1}$, $J$, $Y$ and $\nu$
replaced by $J_0$, $Y_0$ and $\nu_0$, and $z$ replaced by $\ttz =
q_0 \tte$. All we need to obtain the complete description of the
perturbation is to match expressions for $\B_0$, $\B_s$ and $\B_1$
with the help of expressions for $\E_0$ and $\E_s$ at the moments
$\eta_s$ and $\eta_{rad}$.

At the moment $\eta_s$, the jumps in $w_\eta$ and $\b_\eta$ are
$\D w_s = w - w_0 \equiv \D w$ and $\D \b_s = \b$. By using these
values and the identity $\E_0 = -3\H_s \B_{0s}$ we find
\begin{equation*}
[\B]_s = 0, \quad [\B']_s = - \Bl \frac 12 \D w - \frac 43 \b \Br
\E_0,
\end{equation*}
The resulting equations for the unknowns $\tx = c_J J \tz_s^{-m}$
and $\ty = c_Y Y \tz_s^{-M}$ are
\begin{equation}
\tx + \ty = C \frac {u_0}u, \quad (1 - m) \tx + (1 - M) \ty = C
\BL 1 - \Bl \b - \frac 38 \D w \Br 8u_0 \BR,
 \label{eq:tmatch}
\end{equation}
and their solution is
\begin{equation}
\tx = C \frac {u_0}u \frac 1{2n} (M - \s_2), \quad \ty = - C \frac
{u_0}u \frac 1{2n} (m - \s_2).
 \label{eq:tsolxy}
\end{equation}

Potentials $\Phi$ and $\Psi$, computed from the functions $\B$ and
$\E$, are enhanced by a factor of order $\b \tz_s^{-2} \Pz$ for
$\eta$ close to $\eta_s$. The former potential jumps either up or
down by the value of that order, while the latter potential rises
or falls abruptly without a jump. (This follows from the fact that
$\tx$ and $\ty$ are proportional to $M - \s_2$ and $- (m - \s_2)$,
just as $x$ and $y$ computed earlier for a universe in which the
parameter $w$ did not change during the fluid-to-solid
transition.) As before, exploding $\Phi$ and $\Psi$ do not disrupt
the theory, since $\B$ and $\E$ vary smoothly enough.

At the moment $\eta_{rad}$, the jumps in $w_\eta$ and $\b_\eta$
are $\D w_{rad} = - \D w$ and $\D \b_{rad} = - \b$. By inserting
these values into the expressions for $[\B]$ and $[\B']$ we obtain
\begin{equation*}
[\B]_{rad} =  \frac {\D w}{w_{0+}} \Bl \B_{s,rad} + \frac
{\E_{rad}}{3\H_{rad}} \Br, \quad [\B']_{rad} = -4 \frac {\D
w}{w_{0+}} \H_{rad} \B_{s,rad} - \Bl \frac {5 - 3w_0}{6w_{0+}} \D
w + \frac 43 \b \Br \E_{rad}.
\end{equation*}
($\B_s$ refers to the function $\B$ between the times $\eta_s$ and
$\eta_{rad}$, hence $\B_{s,rad}$ is the limit of that function for
$\eta$ approaching $\eta_{rad}$ from the left. The quantity
$\E_{rad}$ is to be understood in the same way.) Introduce the
variables
\begin{equation}
\tX = c_J J \tz_{rad}^{-m} = p^{-m} \tx, \quad \tY = c_Y Y
\tz_{rad}^{-M} = p^{-M} \ty,
 \label{eq:tXtY}
\end{equation}
where $p$ is the ratio of final and initial moments of the period
during which the solid affects the dynamics of the universe, $p =
\te_{rad}/\te_s$. Equations for the unknowns $\ttx = c_{J1} J_0$
and $\tty = c_{Y1} Y_0 \ttz_{rad}^{-2\nu_0}$ are
\begin{equation}
\ttx + \tty = \frac q{q_0} \frac u{u_0} (K_J\tX + K_Y\tY), \quad
\ttx + (1 - 2\nu_0) \tty = \frac q{q_0}(L_J\tX + L_Y\tY),
 \label{eq:ttmatch}
\end{equation}
where the coefficients on the right hand side are defined as
\begin{equation*}
K_J = \frac 1{w_{0+}} \BL w_+ - \frac {\D w}{6u\tw} (m + \tau)
\BR, \quad K_Y = \mbox{ditto with } m \to M,
\end{equation*}
and
\begin{equation*}
L_J = 1 - m - \frac {8u\D w}{w_{0+}} + \frac {m + \tau}{\tw} \Bl
\frac{5 - 3w_0}{6w_{0+}} \D w + \frac 43 \b\Br, \quad L_Y =
\mbox{ditto with } m \to M,
\end{equation*}
The solution is
\begin{equation}
\ttx = \frac 1{2\nu_0} \frac q{q_0} (M_J \tX + M_Y \tY). \quad
\tty = - \frac 1{2\nu_0} \frac q{q_0} (N_J \tX + N_Y \tY)
 \label{eq:ttxy}
\end{equation}
with the constants $M_\a$ and $N_\a$, $\a = J$, $Y$, defined in
terms of the constants $L_\a$ and $K_\a$ as
\begin{equation*}
M_\a = L_\a - (1 - 2\nu_0) \frac u{u_0} K_\a, \quad N_\a = L_\a -
\frac u{u_0} K_\a.
\end{equation*}

The nondecaying part of the function $\Phi$ in the period after
the radiation takes over again is
\begin{equation*}
\Phi_1 = C_{J1} J_{0+} = \frac {\S_0 q_0}{2\nu_{0+}} \ttx.
\end{equation*}
Here we must insert for $\ttx$ from equation (\ref{eq:ttxy}), with
$\tX$ and $\tY$ given in equation (\ref{eq:tXtY}) and $\tx$ and
$\ty$ given in equation (\ref{eq:tsolxy}). The constant $C$ in the
latter equation is given by the expression following from equation
(\ref{eq:P0}) with $\S$ replaced by $\S_0$ and $\nu$ replaced by
$\nu_0$,
\begin{equation*}
C = \frac {2\nu_{0+}}{\S_0 q} \Pz.
\end{equation*}
The resulting expression for $\Phi_1$ is
\begin{equation}
\Phi_1 = \frac 1{2\nu_0} \frac {u_0}u \frac 1{2n} (\hM_J p^{-m} -
\hM_Y p^{-M}) \Pz,
 \label{eq:P1}
\end{equation}
with the coefficients $\hM_J$ and $\hM_Y$ defined as
\begin{equation*}
\hM_J = M_J (M - \s_2), \quad \hM_Y = M_Y (m - \s_2).
\end{equation*}
After some algebra the coefficients reduce to
\begin{equation}
\hM_J = 2\nu_0 \frac u{u_0} M - b, \quad \hM_Y = \mbox{ditto with
} M \to m.
 \label{eq:MJMY}
\end{equation}

In a universe filled with ideal fluid, the potentials $\Phi$ and
$\Psi$ coincide and are continuous together with their derivative
at the moment when $w$ jumps to the new value. Thus, if $\Phi$ did
not contain the decaying term at the beginning, it does not
develop it during the jump. As a result, its value stays the same.
(This is not true if $w$ changes continuously. For example, during
the radiation-to-matter transition shortly before recombination
$\Phi$ decreases by the factor 9/10.) In our problem with $\b =
0$, the medium filling the universe is ideal fluid whose parameter
$w$ changes abruptly from one value to another and back again,
therefore $\Phi_1$ (final value of $\Phi$) equals $\Pz$ (initial
value of $\Phi$). The same result is obtained from equations
(\ref{eq:P1}) and (\ref{eq:MJMY}), if we insert $n = \nu$, $M =
2\nu$ and $m = b = 0$ into them.

Let us now determine how fast the function $\Phi$ approaches its
limit value. The decaying part of $\Phi$ in the period under
consideration is
\begin{equation}
\D \Phi_1 = -2\nu_{0+} \frac {u_0}u \frac 1{2n} (\hN_J p^{-m} -
\hN_Y p^{-M}) \ttz_{rad}^{-2} \z^{-2 \nu_{0+}} \Pz,
 \label{eq:DP1}
\end{equation}
where $\z$ is rescaled time normalized to 1 at the moment
$\eta_{rad}$, $\z = \ttz/\ttz_{rad}$, and the coefficients $\hN_J$
and $\hN_Y$ are defined in terms of $N_J$ and $N_Y$ in the same
way as the coefficients $\hM_J$ and $\hM_Y$ in terms of $M_J$ and
$M_Y$. After rewriting the former coefficients similarly as we did
with the latter ones, we obtain
\begin{equation}
\hN_J = \hN_Y = -\frac {w_0}{w_{0+}} 2b.
 \label{eq:NJNY}
\end{equation}
From these equations and equations (\ref{eq:P1}) and
(\ref{eq:MJMY}) we find that the ratio of the decaying and
nondecaying part of $\Psi$ at the moment of solid-to-radiation
transition is
\begin{equation}
\frac {\Delta \Phi}{\Phi_1}\Big |_{rad} = R_{rad} \ttz_{rad}^{-2},
\quad R_{rad} = 4\nu_{0} \nu_{0+} \frac {w_0}{w_{0+}}
\frac{2u_0b}{2\nu_0 u [n \coth (n \log p) + \nu] - u_0 b}.
 \label{eq:Rrad}
\end{equation}
The ratio is greater than one for $|\b| \gtrsim \ttz_{rad}^2$. The
function $\Phi$ is then dominated by the decaying term at the
moment $\eta_{rad}$, and the nondecaying term takes over later, at
the moment $\eta_{nd}$ given by
\begin{equation}
\ttz_{nd} = R_{rad}^{\frac 1{2\nu_{0+}}} \ttz_{rad}^{1 - \frac
1{\nu_{0+}}}.
 \label{eq:zdec}
\end{equation}
The exponent at $\ttz_{rad}$ is positive for any $w_0 < 1$ (it
equals 1/3 for $w_0 = 1/3$) and the constant $ R_{rad}$ is of
order 1 or less. Thus, if the perturbation was large-scale at the
moment the fluid originally filling the universe started to be
dominating again ($\ttz_{rad} \ll 1$), it will be still
large-scale at the moment the nondecaying term prevails over the
decaying one ($\ttz_{nd} \ll 1$).

The time $\eta_{rad}$ must not be too close to the time of
recombination, if the spectrum of large-size CMB anisotropies is
not to be distorted. Suppose $w_0 = 1/3$ and denote the values of
the field $\Phi_{1tot} = \Phi_1 + \D \Phi$ which it assumes at the
moment $\eta_{re}$ for wave numbers $k^\0$ and $0.01 k^\0$ by
$\Phi^\0$ and $\Phi^\1$. Their ratio is
\begin{equation*}
\frac {\Phi^\0}{\Phi^\1} = \frac {1 + R_{rad} \ttz_{rad}^\0}{1 +
10^4 R_{rad} \ttz_{rad}^\0} \doteq 1 - 10^4 R_{rad} \ttz_{rad}^\0.
\end{equation*}
The expression on the left hand side equals 0.01$^{n_S - 1}$,
where $n_S$ is the spectral index, whose deviation from 1 (about
$-0,04$ according to observations) describes the tilt of the
scalar spectrum. If we allow for a tilt of the primordial
spectrum, too, the right hand side will be multiplied by
0.01$^{n_{S0} - 1}$. Denote $p_* = 1/\ttz_{rad}^\0 =
\tte_{re}/\tte_{rad} = a_{re}/a_{rad} = T_{rad}/T_{re}$ and
require that $n_{S0}$ differs from $n_S$ at most by some $\D n_S
\ll 1$. To ensure that, $p_*$ must satisfy
\begin{equation}
p_* > 2 \times 10^3 R_{rad} \D n_S^{-1}.
\end{equation}

For numerical calculations we need the value of $p$. It is a ratio
of {\it times}, but can be rewritten in terms of a ratio of {\it
scale parameters} or {\it temperatures}, $P = a_{rad}/a_s =
T_s/T_{rad}$, as
\begin{equation}
p = P^{\frac 1{2u}}.
 \label{eq:pP}
\end{equation}
The value of $p$, or equivalently, $P$, determines the interval of
admissible $w$'s. To obtain it, note that for $w_0 = 1/3$ equation
(\ref{eq:aeq}) yields $P = (\e^{-1} - 1)^{1/\D} \doteq
\e^{-1/\D}$, or
\begin{equation}
P \doteq \e^{- \frac 1{3\D w}}.
 \label{eq:Papp}
\end{equation}
(This is consistent with equation (\ref{eq:eeq}), which can be
rewritten as $p = \e^{- \frac {\D/2 + 1}\D} = \e^{- \frac 1{6u\D
w}}$.) Thus, the jump in the parameter $w$ for the given ratio $P$
must satisfy
\begin{equation}
\D w \doteq \frac {\log 1/\e}{3 \log P} \gtrsim \frac 1{3\log P}.
 \label{eq:Dw}
\end{equation}

The dependence of the quantities $\tP_1 = \Phi_1/\Pz$ and
$R_{rad}$ on the parameter $\b$ is depicted in fig. 2.
\begin{figure}[ht]
\centerline{\includegraphics[width=0.85\textwidth]{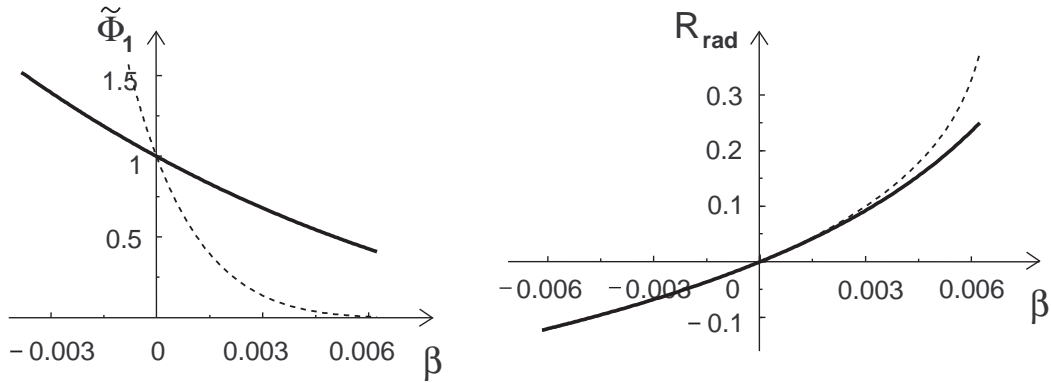}}
\centerline{\parbox {12.5cm}{\caption{\small Final value of
Newtonian potential in a universe with stiff solid (left) and
normalized ratio of decaying to nondecaying part of the potential
at solid-to-radiation transition (right), plotted as functions of
shear modulus}}}
  \label{fig:PP0}
\end{figure}
The values of $w_0$ and $w$ are 1/3 and 2/3 on both panels, and
the solid and dotted lines correspond to $P = 10^3$ and $P =
10^{13}$ respectively. The lines are terminated at $\b = 1/160$,
which is the maximum admissible $\b$ for $w = 2/3$.

The parameter $P$ assumes the smaller value if, for example, the
solid dominated the dynamics of the universe between the
electroweak and confinement scale, and the greater value, if the
solid was formed as soon as at the GUT scale and dominated the
dynamics of the universe up to the electroweak scale. Unless the
parameter $w$ of the solid is close to that of radiation, the
fraction of energy which remains stored in radiation after the
solid has been formed must be quite small in the former case and
very small in the latter case. For $w = 2/3$ this fraction equals
$1/P$, so that for the greater $P$ the mechanism of the
radiation-to-solid transition must transfer to the solid all but
one part in 10 trillions of the energy of radiation.

The quantity $\tP_1$ is the factor by which the function $\Phi$
changes due to the presence of stiff solid in the early universe.
From the figure we can see that $\Phi$ is shifted upwards for $\b
< 0$ and downwards for $\b > 0$, and the enhancement factor
decreases monotonically with $\b$, the steeper the larger the
value of $P$. For maximum $\b$ the function $\Phi$ is suppressed
by the factor 0.41 if $P = 10^3$ and by the factor 0.004 if $P =
10^{13}$.

The quantity $R_{rad}$ determines, together with $\D n_S$, the
minimal duration of the period between the moment when radiation
took over again and recombination. From the figure we can see that
for maximum $\b$ and $P = 10^3$ the temperature at the beginning
of this period had to be at least $8 \times 10^3 \D n_S^{-1}
T_{re} \doteq 0.2\ (\D n_S/0.01)^{-1}$ MeV.

Finally, let us compute the quantities $\Phi_1$ and $R_{rad}$ for
maximum $\b$ as functions of $w$. The maximum $\b$ corresponds to
$n = 0$ and equals
\begin{equation}
\b_m = \frac 3{32} \frac {(1 - w)^2}{w_+}.
 \label{eq:bm}
\end{equation}
If we perform the limit $\b \to \b_m$ in the expression for
$\Phi_1$, we obtain
\begin{equation}
\Phi_{1m} = \frac 1{2\nu_0} \frac {u_0}u (\hM_{0m} \log p + \hM_1)
p^{-\nu} \Pz,
 \label{eq:P1m}
\end{equation}
where $\hM_0$ and $\hM_1$ are the first two coefficients in the
expansion of $\hM_J$ in the powers of $n$,
\begin{equation*}
\hM_0 = 2\nu_0 \frac u{u_0} \nu - b, \quad \hM_1 = 2\nu_0 \frac
u{u_0}.
\end{equation*}
The quantity $R_{rad}$ computed in the limit $\b \to \b_m$ is
\begin{equation}
R_{rad,m} = 4\nu_{0} \nu_{0+} \frac {w_0}{w_{0+}}
\frac{2u_0b_m}{2\nu_0 u (1/\log p + \nu) - u_0 b_m}.
 \label{eq:Rm}
\end{equation}
For $w = 1$ it holds $\b_m = 0$ and hence $\Phi_{1m} = \Pz$ and
$R_{rad,m} = 0$. As $w$ decreases, $\Phi_{1m}$ decreases, too, and
it reaches minimum for the value of $\D w$ given on the right hand
side of equation (\ref{eq:Dw}). For $P = 10^{13}$, the minimum is
as small as 3$\tim$10$^{-6}$. The quantity $R_{rad,m}$, on the
other hand, decreases with {\it increasing} $w$; for example, it
falls from 0.82 to 0 for $P = 10^3$ and from 1.28 to 0 for $P =
10^{13}$, as $w$ increases from the minimum value given in
equation (\ref{eq:Dw}) to 1. (All numerical values refer to $w_0 =
1/3$.)


\section{Conclusion}
 \label{sec:con}

We have studied the effect of solid matter with $w > 0$ on the
evolution of scalar perturbations. Two scenarios were analyzed: a
scenario with radiation-like solid appearing in the universe at
some moment during the radiation era and staying there till
recombination, and a scenario with stiff solid appearing also in
the radiation era and dominating the evolution of the universe
during a limited period before recombination. The focus was on
long-wavelength (supercurvature) perturbations, therefore it was
necessary to assume that the solidification was anisotropic,
producing a solid with flat internal geometry; otherwise no new
effect would be obtained. In the calculations, proper-time
comoving gauge was used instead of more common, and intuitively
more appealing, Newtonian gauge. Besides being more convenient
computationally, the gauge turned out to be preferable also on
principal grounds, since the requirement of smallness of
perturbations was not violated in it. (Note that the description
of long-wavelength perturbations in Newtonian gauge breaks down
also in case $w < 0$, because evolutionary equations are then
numerically unstable. This forced the authors of \cite{bs} to
switch to {\it synchronous gauge} after formulating the theory in
Newtonian gauge.) Any theory of long-wavelength perturbations must
be consistent with the observational fact that their spectrum at
recombination is flat. We have shown that in the problem with
radiation-like solid this leads to a constraint on $\xi$
(dimensionless shear modulus), and in the problem with stiff solid
this yields a constraint on the ratio of $T_{rad}$ (temperature at
solid-to-radiation transition) to $T_{re}$ (temperature at
recombination): in order that the theory does not contradict
observations, $|\xi|$ must be small enough and $T_{rad}/T_{re}$
must be large enough. The net effect of stiff solid is suppression
of Newtonian potential in case $\xi > 0$ and enhancement of it in
case $\xi < 0$. This might raise hope that for $\xi > 0$ also the
scalar-to-tensor ratio is suppressed, which would surely be an
interesting effect from the observational point of view. However,
a straightforward calculation shows that tensor perturbations are
suppressed by exactly the same factor as scalar ones. Thus, the
ratio would change only if its {\it initial value} would change;
and whether this happens or not depends on how the inflationary
scenario is modified to provide appropriate initial conditions for
scalar perturbations.


\appendix
\section{Relasticity}
 \label{app:rel} \setcounter{equation}{0}
\renewcommand{\theequation}{A-\arabic{equation}}

Consider an elastic medium put into the metric $g_{\mu \nu}$,
whose body coordinates $X^A$ are given functions of the spacetime
coordinates $x^\mu$. The deformation of the medium is described by
the body metric $H^{AB}$, defined as the spacetime metric
push-forwarded to the body space,
\begin{equation}
H^{AB} = - g^{\mu \nu} {X^A}_{,\mu} {X^B}_{,\nu}.
 \label{eq:H}
\end{equation}
Material properties of the medium are encoded in the constitutive
equation $\r = \r(H^{AB})$. Knowing the function $\r(H^{AB})$, one
computes the energy-momentum tensor as
\begin{equation}
T_{\mu \nu} = 2 \frac {\partial \r}{\partial H^{AB}} {X^A}_{,\mu}
{X^B}_{,\nu} + \r g_{\mu \nu}.
  \label{eq:Ts0}
\end{equation}

A special kind of elastic medium is an ideal fluid. To define it,
one introduces particle density $n$. In general, $n$ is
proportional to the particle density $n_{ref}$ which would be
observed in the medium if transformed into some properly chosen
reference relaxed state. If the space is filled with one kind of
medium only, the density $n_{ref}$ can be rescaled to 1 and the
actual particle density can be written as
\begin{equation}
n = (\det H^{AB})^{1/2}.
   \label{eq:n}
\end{equation}
By definition, ideal fluid is a medium whose energy density
depends on $H^{AB}$ only through $n$.

An important new concept in relasticity is that of the {\it
partially relaxed state}, defined as the state in which the medium
has minimum energy per particle $\e = \r/n$ at fixed $n$. Consider
a state close to the partially relaxed state and write the
quantities appearing in equations (\ref{eq:H}), (\ref{eq:Ts0}) and
(\ref{eq:n}) as $f = f^\0 + \d f$, where $f^\0$ is the value in
the partially relaxed state and $\d f$ is a small corrections to
it. If the medium is isotropic, the constitutive equation reads
\begin{equation}
\d \e = - \s^\0 \frac{\d V}{V^\0} + \frac 18 \bl (\d {H_A}^A)^2 +
\frac 14 \bar \mu (\d {H_A}^B)^2,
  \label{eq:ceq}
\end{equation}
where $\s$ is pressure energy per particle, $\s = p/n$, $\bl$ and
$\bar \mu$ are Lame coefficients per particle, $\bl = \l/n$ and
$\bar \mu = \mu/n$, $V$ is volume per particle, $V = 1/n$, and the
first index of the tensor $\d H^{AB}$ is lowered by the matrix
$H^\0_{AB}$ inverse to the matrix $H^{\0 AB}$, $\d {H_A}^B =
H^\0_{AC} \d H^{CB}$. In the last therm, the ``implicit summation
rule'' is used, $(\d {H_A}^B)^2 = \d {H_A}^B \d {H_B}^A$. To
compute $T_{\mu \nu}$, we need to express $\d \e$ in terms od $\d
{H_A}^B$ only. This is achieved by writing the ratio $\d V/V^\0$
on the right hand side of (\ref{eq:ceq}) as
\begin{equation*}
\frac{\d V}{V^\0} = \frac 12 \d {H_A}^A + \frac 12 (\d {H_A}^A)^2
+ \frac 14 (\d {H_A}^B)^2.
\end{equation*}

Equation (\ref{eq:ceq}) holds to the second order in $\d {H_A}^B$.
Within this accuracy, the trace $\d {H_A}^A$ in the $\l$-term can
be replaced by $2 \d V/V$, so that if $\mu$ vanishes, $\d \e$ as
well as $\d \r$ depends on $\d H$ only through $\d V$ and we are
dealing with an ideal fluid. Note also that by comparing
(\ref{eq:ceq}) to the Taylor expansion of $\e(V^\0 + \d V)$,
passing from $\e$ and $\s$ to $\r$ and $p$ and skipping the index
(0), one obtains
\begin{equation}
\frac{d\r}{dV} = - \frac {\r_+}V, \quad \frac{dp}{dV} = -\frac KV,
 \label{eq:id0}
\end{equation}
where $\r_+$ and $K$ are defined in section \ref{sec:cosmo}. If we
introduce an auxiliary sound speed $c_{S0}$ defined in terms of
the function $p(\r)$ in the same way as the sound speed of an
ideal fluid, $c_{S0}^2 = dp/d\r$, we find
\begin{equation}
c_{S0}^2 = \frac K{\r_+}.
 \label{eq:cS20}
\end{equation}

In an unperturbed universe, the 3-space coordinates are comoving,
${\bf x} = {\bf X}$, and the matter at any given moment is in a
partially relaxed state with $H^{\0 ij} = - g^{ij} = a^{-2}
\d_{ij}$. In a perturbed universe, the 3-space coordinates differ
from the body coordinates by a small displacement vector ${\bm
\xi}$, ${\bf x} = {\bf X} + {\bm \xi}$, and the body metric
acquires a small correction $\d H^{ij}$. This yields
\begin{equation}
{T_0}^0 = \r^\0 - \frac 12 \r_+ \d {H_k}^k, \quad {T_i}^0 = \r_+
(- {\xi^i}' + h_{0i}), \quad {T_i}^j = - p^\0 \d_{ij} + \frac 12
\l \d {H_k}^k \d_{ij} + \mu \d {H_i}^j.
 \label{eq:Ts}
\end{equation}
With the index (0) at $\r$ and $p$ skipped, these expressions
reduce to those cited in section \ref{sec:cosmo}. Let us verify
that. In an interval of conformal time $d\eta$, the proper time
$\tau$ of any given volume element of the medium increases
approximately by $d \tau = a d\eta$, so that ${\bf u} \equiv d{\bf
x}/d\tau = a^{-1} {\bm \xi}'$, $u_i = a(- {\xi^i}' + h_{0i})$ and
expression for ${T_i}^0$ coincides with that in equation
(\ref{eq:Tf}). By comparing the expressions for ${T_0}^0$ and
${T_i}^j$ with those in extended equation (\ref{eq:Tf}) we obtain
\begin{equation}
\d \r = - \frac 12 \r_+ \d {H_k}^k, \quad \d p = - \frac 12 K \d
{H_k}^k, \quad \D {T_i}^j = \mu \d \hH_i^{\ j},
 \label{eq:d}
\end{equation}
where the tilde denotes the traceless part of the matrix, $\d
\hH_i^{\ j} = \d {H_i}^j - \frac 13 \d {H_k}^k \d_{ij}$. From the
definition of $H^{AB}$ it follows
\begin{equation}
\d {H_i}^j = {\xi^i}_{,j} + {\xi^j}_{,i} - h_{ij},
 \label{eq:dH}
\end{equation}
so that in the comoving gauge used throughout the paper, in which
$\xi^i = 0$, we have $\d {H_i}^j = - h_{ij}$. After inserting this
into equation (\ref{eq:d}) and using the definition of $\d
\tau_{ij}$, we arrive at the expressions for $\d \r$ and $\d
\tau_{ij}$ in equation (\ref{eq:dcom}). Note also that from the
first two equations it follows that the ratio $\d p/\d \r$ equals
the derivative $d p/d \r$.

Equation (\ref{eq:dH}) holds only for elastic media with {\it flat
internal geometry}. In case the internal geometry is perturbed, we
must distinguish between local body coordinates $X^i$ and global
body coordinates ${\cal X}^i$, and write the body metric tensor in
the latter coordinates as
\begin{equation*}
\H^{ij} \equiv \frac{\partial {\cal X}^i}{\partial X^k}
\frac{\partial {\cal X}^j}{\partial X^l} H^{kl} = a^{-2} (\d_{ij}
+ b_{ij}).
\end{equation*}
The formulas (\ref{eq:d}) then remain valid, but in the formula
(\ref{eq:dH}) there appears an extra term
\begin{equation}
\D {H_i}^j = - b_{ij}.
 \label{eq:DdH}
\end{equation}

When using the expression (\ref{eq:ceq}) for $\d \e$ in a
perturbed universe, we have tacitly assumed that the entropy per
particle $S$ is constant throughout the space. Thus, we have
considered {\it adiabatic perturbations} only. More general are
{\it entropy perturbations} which include nonzero correction to
the entropy per particle $\d S$. For such perturbations, there
appear additional terms proportional to $\d S$ in the first two
formulas in (\ref{eq:d}). The explicit form of these terms is
\begin{equation}
\D \r = nT\d S, \quad \D p = n^2 \left(\frac{\partial T}{\partial
n}\right)_{\mbox{\hskip -1mm}S}\d S,
 \label{eq:Dd}
\end{equation}
where $T$ is the temperature of the medium.

Consider a perturbation in the form of plane wave with the given
wave vector $\bf k$, and suppose it is located well inside the
horizon, $k \eta \gg 1$. For such perturbation, longitudinal and
transverse sound speeds $c_{S\|}$ and $c_{S\bot}$ are given within
a good accuracy by the same formulas as in Minkowski space,
\begin{equation}
c_{S\|}^2 = \frac {\l + 2\mu}{\r_+}, \quad c_{S\bot}^2 = \frac
\mu{\r_+}.
 \label{eq:cS2}
\end{equation}
Corrections to these expressions are of order $(k\eta)^{-2}$.

\end{document}